\documentclass[11pt,a4paper]{article}
\usepackage{jheppub}
\usepackage{booktabs}

\usepackage{slashed}
\usepackage{graphicx,amssymb,amsmath,amsfonts}
\usepackage{hyperref}
\usepackage[utf8]{inputenc}

\usepackage[english]{babel}

\newcommand{\pt}[1]{\left(#1\right)}

\newcommand{\pmat}{\begin{pmatrix}}
\newcommand{\fpmat}{\end{pmatrix}}
\newcommand{\eq}{\begin{equation}}
\newcommand{\feq}{\end{equation}}
\newcommand{\cas}{\begin{cases}}
\newcommand{\fcas}{\end{cases}}

\newcommand{\eqarray}{\begin{eqnarray}}
\newcommand{\feqarray}{\end{eqnarray}}
\newcommand{\Tr}[1]{\operatorname{Tr}\pt{#1}}

\newcommand{\Imm}{\,\text{Im}\,}
\newcommand{\Rea}{\,\text{Re}\,}

\newcommand{\alp}{\alpha{}'{}}
\newcommand{\alpe}{\alpha_0^\prime{}}

\newcommand{\be}{\beta}

\newcommand{\m}{\mu}

						\def\m{\mu}

\def\be{\begin{equation}}
\def\ee{\end{equation}}
\def\bea{\begin{eqnarray}}
\def\eea{\end{eqnarray}}
\title{Partonic behavior of string scattering amplitudes
from holographic QCD models}

\author[a,b]{Massimo Bianchi,}
\author[a,b]{Maurizio Firrotta,}
\author[c,d]{Jacob Sonnenschein,}
\author[e]{Dorin Weissman}

\affiliation[a]{Dipartimento di Fisica, Università di Roma Tor Vergata\\
Via della Ricerca Scientifica 1, 00133, Roma, Italy}
\affiliation[b]{INFN sezione di Roma Tor Vergata \\
Via della Ricerca Scientifica 1, 00133 Roma, Italy}
\affiliation[c]{The Raymond and Beverly Sackler School of Physics and Astronomy, Tel Aviv University, \\ Ramat Aviv 69978, Tel Aviv, Israel}
\affiliation[d]{Center for Cosmology and Particle Physics
Department of Physics, New York University,\\ 726 Broadway, New York, NY 10003, USA}
\affiliation[e]{Okinawa Institute of Science and Technology, \\ 
1919-1 Tancha, Onna-son, Okinawa 904-0495, Japan}

\emailAdd{massimo.bianchi@roma2.infn.it}
\emailAdd{mfirrotta@roma2.infn.it}
\emailAdd{cobi@tauex.tau.ac.il}
\emailAdd{dorin.weissman@oist.jp}

\abstract{We study the emergence of partonic behavior in scattering processes at large Mandelstam's variable $s$  from string amplitudes in holographic backgrounds. We generalize the approach of Polchinski and Strassler \cite{Polchinski:2001tt} in two ways. (i) We analyze several holographic confining backgrounds, in particular the hard wall model, the soft wall model and Witten's model. (ii) In addition to deriving the asymptotic behavior of the amplitudes at fixed angle and in the Regge limit,  we also  expand the amplitudes around their poles, integrate over the holographic direction and then re-sum the expansion. Due to dependence of the string tension  on the holographic coordinate, the resulting singularities take the form of branch points rather than poles and the amplitudes display branch cuts and acquire a finite imaginary part. This may signal the failure of the PS  prescription  
 to reproduce the correct analytic structure at low energies. We also observe that the peaks are more pronounced in the region of small $s$ but fade away for large $s$.
In the fixed angle approximation we find in the hard and soft wall models  that
${\cal A}\sim s^{2-\Delta/2}$ whereas in Witten's model ${\cal A} \sim s^{3-\Delta/2}$ and ${\cal A} \sim s^{7/3-2\Delta/3}$ for the 11D and 10D formulations, respectively. In the Regge regime ${\cal A}\sim {s^{2}}\,t^{-2+\alpha}\,(\log{s/ t})^{-1+\alpha}$ where $\alpha$ is the power found in the fixed angle regime.  Using the pole expansion the result for each model is $\Rea[{\cal A}] \sim s^{-1}$, $\Imm[{\cal A}] \sim s^{\alpha}$.
We compute the corresponding amplitudes for mesons using  open strings and find qualitatively similar results as for closed strings.}

\begin{document}
\maketitle

%%%%%%%%%%%%%%%%%%%%%%%%%%%%%%%%%%%%%%%%%%%%

\section{Introduction}
The 1968  deep  inelastic experiment at SLAC provided evidence that in the high energy, fixed angle regime hadronic scattering amplitudes fall off according to a power law behavior in the center of mass energy (or Mandelstam variable $s$). On the other hand Veneziano's and all string amplitudes in this regime display a soft, exponential behavior in $s$ \cite{Veneziano:1968yb, Virasoro:1969me,Green:1982sw}.  This clash between theory and experiment led to abandoning  the description of strong interactions in terms of  string theory.  Yet, string theory has been very successful in describing the low energy spectra of  hadrons.   For hadrons built from light quarks the experimental data that have been accumulated since  the early days of the dual resonance models and up to now fit nicely  Regge-like trajectories where $M^2$ is (almost) linear with the angular momentum $J$ and the excitation number $n$.
For hadrons that carry $s$, $c$ or $b$ flavors, the data fit well a model of a string with massive endpoints \cite{Sonnenschein:2014jwa}.
 
 In terms of Regge trajectories, the passage from ``soft" to ``hard" scattering takes the form of the bending of the trajectory. At positive Mandelstam $t$ we have the linear trajectory  $\alpha(t) = \alp t + ...$, which turns into an almost flat horizontal line for large negative $t$, $\alpha(t) = \mathrm{Const.}$, which characterizes asymptotic freedom.\footnote{A mild logarithmic behavior $\alpha(t)\approx \Lambda^{-2} \log t/\Lambda^2$ would also be acceptable.} The nature of this transition is still an open problem in QCD.

One of the main challenges for a reliable string theory of hadron physics  has been to construct a description that admit spectra on  Regge-like trajectories  and at the same time be  in accordance with the hard  rather than soft behavior of scattering amplitudes at fixed angle.
%In addition one would like  to ``derive" the string theory of hadrons from the underlying QCD theory.

For several decades  this challenge has  not  been successfully addressed  until holography came to rescue. In that  context a
 pioneering work  has been carried out by Polchinski and Strassler \cite{Polchinski:2001tt}, that argued for the hard `partonic' behavior of glueball scattering at large energies and fixed angle in the $AdS$ hard wall model. The original short letter \cite{Polchinski:2001tt} was expanded upon in \cite{Polchinski:2002jw} and \cite{Brower:2006ea}.
%This task was undertaken by J. Polchinski and M. Strassler in \cite{Polchinski:2001ttass} where they showed that indeed in the Ads hard wall model for closed strings the holographic amplitude does turn into a hard scattering amplitude one. 

Their basic proposal is that the four dimensional amplitude can be approximated by
\be {\cal A}_4(s,t,u) \propto \int dr \sqrt{-g} \times {\cal A}_{10}[\tilde s(r), \tilde t(r), \tilde u(r)] \times \prod_{i} \Psi_i(r) \label{eq1} \ee
where \(r\) is the holographic radial dimension, \({\cal A}_{10}\) the amplitude of string theory in 10D flat spacetime, \(\Psi(r)\) the wave functions of the states being scattered, and, crucially, the Mandelstam variables \(\tilde s\), \(\tilde t\), and \(\tilde u\) are rescaled as a function of \(r\). This scaling can also be viewed as a rescaling of the  slope of the trajectories or  the  string tension.

There are two different descriptions of hadrons in holography: (i) As fluctuations of fields that reside either in the bulk or on probe flavor branes. The former are associated with glueballs and the latter with mesons. Baryons are described as flavor instantons on probe branes. (ii) As string configurations in the holographic backgrounds. Closed strings are naturally the duals of glueballs and open strings with endpoints on flavor branes are the duals  of  either mesons or baryons. 
%In figure (\ref{f.strings}) we present cartoons of the different holographic stringy hadrons. 

The spectra of hadrons have been extracted in approach (i) by writing down the wave equations of the various different modes, diagonalizing the equations  and determining the resulting eigenvalues. An intensive research effort has been devoted to this task. For a review of this direction see \cite{Erdmenger:2007cm}. It is probably fair to state that, apart from one exception, the outcome spectra do not fit well the experimental data. This is manifested simply  by the fact that they do not admit a Regge-like behavior, but rather a Kaluza-Klein-type spectrum $m_n^2\sim n^2/R^2$. The exception to that is the soft wall model \cite{Karch:2006pv} where  Regge behavior does show up, but only with respect to the excitation number $n$ and not the angular momentum $J$.

As for the spectra in the stringy approach, for strings in flat spacetime the Regge-like spectrum is an almost trivial result. In the higher dimensional curved holographic backgrounds it is a difficult task to determine the quantum string configurations and hence the spectra were not determined in this way. One can perform semiclassical computations, as e.g. in \cite{Frolov:2002av}. In \cite{Sonnenschein:2016pim} the holography inspired stringy hadron model (HISH) was proposed based on the observation of \cite{PandoZayas:2003yb} that showed that the classical solutions of the string equations of motion in holographic confining backgrounds can be mapped to those of strings in flat four dimensions but with massive particles on their endpoints. Based on this map semiclassical spectra  of hadronic strings have been theoretically computed and compared with experimental data. Explicit fits of the data and the HISH model were done in \cite{Sonnenschein:2014jwa} for mesons, and  \cite{Sonnenschein:2014bia} for baryons. The model was also extended to exotic hadrons and many of its predictions are reviewed in \cite{Sonnenschein:2018fph}.
%This nice correspondence applies  in particular   also to  the stringy hadrons picture used in the   of the work of \cite{Polchinski:2001tt}.  
%Using the map to HISH, it is thus clear that indeed the spectra determined by  the stringy holographic approach (ii) 

%The nice agreement between the measured spectra and that associated with the stringy hologrphic picture of hadrons does not occur for the spectra computed for hadrons as holographic fields, with the exception of the 

To calculate string amplitudes in holographic  settings  one should generalize the procedure for the  derivation of Veneziano amplitude in a 10D (highly) curved spacetime, in $AdS$ or other backgrounds. This is a formidable task. 
Instead, the method proposed in \cite{Polchinski:2001tt} to compute the amplitudes is a ``hybrid" method. One  is instructed to use the amplitudes of  scattering strings in flat spacetime but also dress them with the wave functions associated with the hadronic modes of approach (i) as weights in an integral over the holographic direction. 
 
In this paper we follow this hybrid  formulation  of the amplitude of \cite{Polchinski:2001tt} and show that the transition from soft to hard, or what is referred to as  the bending, holds also in other confining holographic backgrounds. In particular we computed the amplitudes of closed strings for the hard wall model, soft wall model \cite{Karch:2006pv} and  Witten's model \cite{Witten:1998qj}.  We also conjecture that this transition probably exists  in any confining  background.  We then  generalize and extend the analysis to open strings which are the duals of  mesons and baryons in the hard wall, soft wall and the Witten-Sakai-Sugimoto model \cite{Sakai:2004cn}. The case of open strings is complicated by the fact that open strings should be attached to flavor branes, and in this paper we only begin the analysis of these systems.

We find that, while all models admit power law behavior at fixed angle, high energy limit, the value of the power that one obtains is model dependent, and varies  between them. In four dimensional QCD there is a simple expectation from dimensional analysis due to Brodsky and Farrar \cite{Brodsky:1973kr, Brodsky:1974vy},
\be {\cal A} \sim s^{2-\frac{N_p}{2}} \ee
where \(N_p\) is the total number of incoming and outgoing partons of the process. For instance, for \(2\to 2\) meson scattering, \(N_p = 4\times 2 = 8\), and therefore \({\cal A} \sim s^{-2}\). In \cite{Polchinski:2001tt} it was found that in hard wall \(AdS_5\) models, the amplitude behaves as
\be {\cal A} \sim s^{2-\frac{\Delta}{2}} \ee
where now \(\Delta\) is the sum of scaling dimensions of the relevant operators, that contribute as \(\prod_i\psi_i(r) \sim r^{-\Delta}\) in the UV region. Since we want it to be equal to the number of partons to match with QCD, the commonly accepted argument is that one needs to replace the scaling dimension with the twist, as stated already in \cite{Polchinski:2001tt}.

%%%%%%%%%%%%%%%%%%%%%%%%%%%%%%%
In addition to showing the transition from soft to hard scattering, the holographic prescription for the scattering amplitude reveals also the following phenomena: (i)  The ordinary poles of the amplitude of flat spacetime transform into logarithmic  branch points. Starting from these points the amplitude has branch cuts and a finite imaginary part even for real Regge trajectories. This is an immediate consequence of the integration over the radial coordinate and of the scaling of $\alp$ with the holographic coordinate. It seems to be a generic result that would hold also for other holographic confining backgrounds.

(ii) Furthermore, the peaks are more pronounced in the region of small $s$ but fade away for large $s$, when we increase \(s\) and keep the angle fixed, such that $|t|$ is also increased. A possible interpretation to this behavior is that it is an effect of asymptotic freedom: approaching asymptotic behavior the corresponding gauge coupling becomes weaker and hence the binding energy of the bound states becomes smaller.

The results are suggestive, but one could argue that the presence of logarithmic singularities rather than poles is indicative that the Polchinski-Strassler approximation fails to reproduce the correct analytic structure at low energies. This question has to be further investigated.
%%%%%%%%%%%%%%%%%%%%%%%%%%%%%%

%We further analyze  and discuss the fate of the Regge poles after `smearing' resulting from integration in the radial/holographic direction. We  also check the  compatibility of our scattering amplitudes with unitarity, in particular factorization and the  optical theorem. We  also try and argue that 'discrete' holography models \cite{Andreev:2004sy} seem to have troubles with factorization, as already mentioned in \cite{Bianchi:2020cfc}. The generalization to any confining background and the interpretation  in terms of 4d HISH model is also briefly discussed.

Proposals for  effective bending of the trajectories in the negative $t$ region not based on holography were made in the past. In particular  in \cite{Veneziano:2017cks} a model that generalizes the old Lovelace-Shapiro model \cite{Lovelace:1968kjy, Shapiro:1969km} as a toy model for $\pi\pi$ scattering was proposed. The model  admits asymptotically linear and parallel Regge trajectories at positive $t$,  and an effective bending of the trajectories in the negative $t$ region.  The model is based on adding to the usual string-theoretic
amplitudes with universal slope $\alp$ a series of amplitudes with fractional slopes $\frac{\alpha'}{k}$. A similar earlier proposal had been made in \cite{Andreev:2004sy}.  The holographic counterpart of this approach would correspond to a `weight' function
$W(r) = \sqrt{|g|} \prod_i \psi_i(r)$ given by a series of $\delta$-functions i.e. $W(r)=\sum_k w_k \delta(r-r_k)$, with $w_k$ chosen or possibly fixed in such a way as to satisfy unitarity constraints and crossing symmetry. A general analysis of the $S$-matrix of weakly coupled theories of massive higher-spin particles using unitarity and crossing symmetry appeared in \cite{Caron-Huot:2016icg} and universal corrections in \cite{Sever:2017ylk}.

Relations between the  partonic nature of scattering of hadrons and holography were addressed in other papers such as \cite{Kang:2004jd,Hatta:2007he,Pire:2008zf,Albacete:2008ze,Penedones:2007ns,BallonBayona:2007qr,Cornalba:2008sp}.
 	
The plan of the paper is as follows. In section \ref{sec:models} we give a brief review of the different approaches to holographic QCD we utilize in this work: the hard wall, soft wall, and Witten models, as well as a general parametrization of the relevant background fields. We include also a short review of how flavor branes are added to the models to represent quarks.

In section \ref{sec:Polchinski_Strassler} we review the main results of \cite{Polchinski:2001tt}. The main calculations are carried out in section \ref{sec:closed}. First we recall the Virasoro-Shapiro amplitude for 4 dilatons and its various limits, and state the prescription for integrating over the radial dimension to obtain the holographic amplitude in three different methods: pole expansion, fixed angle limit, and Regge limit. The calculations are then done explicitly in the hard wall, soft wall, and Witten models. We also give a more generic parametrization of the background and a general result in section \ref{sec:general_background_closed}. Section \ref{sec:open} repeats the same calculations for mesons using open string amplitudes. It is more concise than section \ref{sec:closed}, since many of the results are essentially the same.

The main results of our analysis are then summarized in section \ref{sec:summary}, which contains our conclusions and indicates lines for further investigation.

%In the appendices we give some technical details of our computations. 

%%%%%%%%%%%%%%%%%%%%%%%%%%%%%%%%%%%%%%%%%%%%

\section{Different approaches to holographic QCD} \label{sec:models}
The original duality between ${\cal N}=4$ SYM theory and string theory in $AdS_5\times S^5$ represents the best established instance of the holographic AdS/CFT correspondence \cite{Aharony:1999ti}. One of several reasons why ${\cal N}=4$ SYM is not the right framework to describe hadrons is that the spectrum is continuous and not discrete as in QCD. 
Indeed starting from the \((d+1)\)-dimensional $AdS$ metric
\be ds^2 = \frac{R^2}{z^2}(\eta_{\mu\nu}dx^\mu dx^\mu + dz^2) \ee
with \(\mu,\nu = 0,1,\ldots d\). The wave equation for a scalar particle of mass \(M\) in the bulk is
\be (\Box + M^2)\Phi = 0 \quad \Rightarrow \quad \eta^{\mu\nu}\partial_\mu\partial_\nu \Phi + \partial_z^2\Phi + \frac{1-d}{z}\partial_z \Phi - \frac{M^2 R^2}{z^2}\Phi = 0 \ee
Taking
\be \Phi(x^\mu,z) = e^{i\eta_{\mu\nu} k^\mu x^\nu} \psi_k(z) \ee
and defining \(m \equiv \sqrt{-k^2} \geq 0\), the solution of the equation is
\be \psi_k(z) = z^{\frac{d}{2}} J_{\Delta-\frac{d}{2}}(m z) \label{eq:psi_Ads} \ee 
where \(J_\alpha\) is the Bessel function, and
\be \Delta = \frac{d}{2} + \sqrt{\frac{d^2}4+M^2 R^2} \label{eq:dimAdS} \ee
or equivalently \(M^2 R^2 = \Delta(\Delta-d)\). The four dimensional mass \(m^2 = -k^2\) is continuous and hence there is no mass gap as expected on the basis of conformal invariance. The scaling dimension \(\Delta\) gives the behavior of the wave function near the boundary, i.e. at small \(z\), \(\psi_k(z) \sim z^\Delta\). It is also equal to the dimension of the dual operator in the conformal boundary theory.\footnote{To be exact there are two allowed behaviors at the UV boundary $\psi(z) \sim z^{d-\Delta}$ and $\psi(z) \sim z^\Delta$. In holographic renormalization \cite{Bianchi:2001de, Bianchi:2001kw} the former corresponds to turning on a source i.e. an operator perturbation $j(x)$, the latter to giving a VEV $v(x)$ to the operator in the boundary CFT.}

\subsection{Hard wall model}
The most basic holographic QCD model is the ``hard wall" (HW) model where the five-dimensional $AdS_5$ space is truncated at a certain value of the radial direction $z=z_0$. The near boundary UV region $z=0$ of the $AdS_5$ space corresponds to the asymptotic free UV region of the gauge theory, while confinement is achieved by the IR hard wall at $z=z_0$ which sets the mass scale for the theory.

The wave function \eqref{eq:psi_Ads} must now satisfy a boundary condition at \(z = z_0\). Imposing boundary conditions will lead to a discrete spectrum of \(m^2\). 

Due to the somewhat arbitrary nature of the hard wall model there is some ambiguity in the choice of boundary condition. The wave function will be finite and normalizable for any $m^2$ as in ordinary $AdS$, so one needs a stronger requirement than regularity or normalizability of the wave function. A standard choice \cite{Zou:2018eam} is to require that the wave function vanishes on the wall (Dirichlet boundary conditions), \(\psi_k(z_0) = 0\), in which case the spectrum is determined by the zeroes of the appropriate Bessel function.

Let us set from now on \(d = 4\) and \(M^2 = 0\), the bulk field is a massless scalar, like the dilaton, in \(AdS_5\). Then, the wave function is
\be \psi_{k}(z) = z^2 J_2 (m z) \label{eq:psi_HW}\ee
and the spectrum after imposing Dirichlet boundary conditions is given by
\be J_2(m z_0) = 0 \qquad \Rightarrow \qquad m z_0 = 5.136,\,\, 8.417,\,\, 11.620,\,\, 14.796\ldots \ee
Another consistent choice of boundary conditions is that of Neumann boundary conditions, \(\partial_z \psi(z_0) = 0\).\footnote{This is the choice of boundary conditions made in \cite{Polchinski:2002jw}.} Since classically the closed string is expected to reside near the wall, it is more natural to have a wave function that is peaked at the wall instead of vanishing there. In this case the spectrum is determined by the zeroes of the Bessel function of a lower order as
\be J_1(m z_0) = 0 \qquad \Rightarrow \qquad m z_0 = 3.832,\,\, 7.016,\,\, 10.173,\,\, 13.324,\,\, \ldots \ee

In both cases the spectrum is approximately linear in the excitation number, \(m_n \sim n\).
\be m^{(\nu)}_n =  \frac{z^{(\nu)}_n}{z_0} \quad : \quad J_\nu(z^{(\nu)}_n) =0
\ee
with $\nu = \Delta -\frac d2$ for Dirichlet, or $\nu = \Delta -\frac d2-1$ for Neumann boundary conditions. For large $n$, we have $z^{(\nu)}_n \approx \left(n+\frac\nu2-\frac14\right)\pi+{\cal O}(1/n)$, thus producing a KK-like spectrum $m_n^2 \sim n^2$ rather than linear Regge trajectories $m_n^2 \sim n$.
 
The wave function \eqref{eq:psi_HW} is the one we use to compute scattering amplitudes in the HW model. We will show in section \ref{sec:Bessel} that the choice of boundary condition does not affect the high energy behavior of the amplitude, but it has consequences for its structure at low energy.

\subsection{Soft wall model} \label{sec:softwall}
Since the HW  model does not admit Regge behavior for the corresponding hadron spectra there was a proposal \cite{Karch:2006pv}  to improve the model by ``softening" the hard wall into a soft wall (SW) model. It was argued that in this way one finds an excitation spectrum of the form $m^2_n\sim n$.

To realize the soft wall, one includes a `dilaton' ${D}=\lambda^2 z^2$ so that $e^{-{D}}$ effectively cuts off the $AdS$ horizon and produces an effective potential of the form
 \be
V_{\text{soft}}(z) = V_{AdS}(z) +\delta V_{\text{soft}}(z)=  \frac{4\mu^2L^2 + \kappa(\kappa+2)}{4z^2} + \lambda^4z^2 + (\kappa-1)^2\lambda^2
\ee
Setting $\ell^2 = \mu^2L^2 + \frac{(\kappa+1)^2}{4} = \nu^2$, the eigenfunctions and eigenvalues are, respectively
\be
\psi_{n,\ell}(z) = z^{\ell + \frac{d}{2}} L_{n,\ell}(\lambda^2z^2) \label{eq:psi_SW}
\ee
\be
m^2_{n,\ell}= 2\lambda^2 (E_{n,\ell}+1) = 2\lambda^2 (2n+\ell+2)
\ee
The eigenfunctions are defined in terms of generalized Laguerre polynomials,
\be
 L_{n,\ell}(x)=\sum_{j=0}^{n}(-1)^{j} \begin{pmatrix}n+\ell \\ n-j \end{pmatrix} \frac{x^{j}}{j!} \label{eq:Laguerre}
\ee

Let us note that, notwithstanding the presence of a non-trivial `dilaton' profile, the metric of the SW model is taken to be $AdS$ very much as in the HW model. This will prove crucial in later computations of scattering amplitudes.
 
 %%%
%%%%%%%%%%%%%%%%%%%%%%%%%%%%%%%%%%%%%%%%%%%%

\subsection{Witten's model} \label{sec:Witten_model}
%An important class of {\it bona fide} (supergravity) models for `holographic QCD' is achieved by compactifing  higher dimensional theories
 %to four dimensions in a way as to (partially) break supersymmetry. A prototype model of this approach is that of a compactified D5 or  NS5 brane on $S^2$ \cite{MaldaNunez??}. In an appropriate decoupling limit this provides a dual of  the $3 + 1$ dimensional ${\cal N}=1$ SYM theory\cite{Maldacena:2000yy}. 
A prototypical top-down model of confining background is the so called Witten's model \cite{Witten:1998zw}. Originally this model was derived by twice compactifying a space coordinate in the background of the near horizon limit of large number of M5 branes, namely $AdS_7\times S^4$ background. It was later realized also by  compactification of one spatial coordinate  of a D4-brane on a circle \cite{Brandhuber:1998er}. The sub-manifold spanned by the compactified coordinate and the radial coordinate has the geometry of a cigar.  Imposing on the circle  anti-periodic boundary conditions for the fermions, in particular the gauginos, renders them  massive and hence supersymmetry is broken. In the limit of small compactified radius one finds that the dual field theory is YM theory in four dimensions, contaminated by KK modes even in the low energy effective description.  

The metric is
\be
ds^2  = \bigg(\frac{U}{L}\bigg)^{\frac32}\big(\eta_{\mu\nu}dx^\mu dx^\nu + f(U)dx_4^2\big) + \bigg(\frac{L}{U}\bigg)^{\frac32}\big(\frac{dU^2}{f(U)} + U^2 d\Omega_4^2\big)  \label{eq:backgroundWitten}
\ee
where $L^3=\pi g_s N_c \ell_s^3$,  and
\be 
f(U)=1- \frac{U_\Lambda^3}{U^3}\ee
The background fields also include the dilaton and 4-form given by
\be
e^\phi = g_s \bigg(\frac{U}{L}\bigg)^{\frac34}\,, \qquad F_4 = \frac{2\pi N_c}{V_4} \epsilon_4
\ee
The coordinate $x_4$ describes a circle of radius $R = \frac{3 U_\Lambda^{1/2}}{2\pi L^{3/2}}$ that sets the mass scale $M_{gb} =1/R$ and $T_{st}= \frac{3 U_\Lambda^{3/2}}{2\pi \ell_s^2 L^{3/2}}$. 

The M-theory embedding of Witten's model is represented by the \(AdS_7\times S_4\) black hole:
\be ds^2 = \frac{r^2}{L^2}\left(\eta_{\mu\nu}dx^\mu dx^\nu + f(r) dx_4^2 + dx_{11}^2\right) + \frac{L^2}{r^2} \frac{1}{f(r)}dr^2 + \frac{L^2}{4} d\Omega_4^2 \label{eq:AdS7}\ee
with
\be f(r) = 1 - \frac{r_{0}^6}{r^6} \ee
The wave equations for the different fluctuation modes in this model were first derived in \cite{Brower:2000rp}.\footnote{A useful reference here is \cite{Brunner:2015oqa}, where one can find the explicit relations between the fluctuations modes of the metric around the 11D background of \eqref{eq:AdS7} to those of the fields in the 10D background of \eqref{eq:backgroundWitten}.} We will consider only the lightest scalar mode, which satisfies the wave equation
\be \frac{1}{r^3}\frac{d}{dr}\bigg(r (r^6-r_{0}^6) \psi^\prime(r)\bigg) + \bigg(m^2L^4 + \frac{432 r^2 r_{0}^{12}}{(5r^6-2r_{0}^6)^2}\bigg)\psi(r) = 0 \label{eq:psi_Wit}\ee 
Far from \(r_0\), the geometry is simply that of \(AdS_7\), and the wave function at large \(r\) behaves as 
\be \psi(r) \approx \frac{1}{r^3}J_3\big(\frac{\mu r_0^3}{r^3}\big) \sim \frac{1}{r^6} \label{eq:dimWit}\ee
as in \eqref{eq:psi_Ads}, where we define 
\be \mu^2 = m^2 \frac{L^4}{r_0^2} \ee
such that \(\mu\) is the four dimensional mass in units of \(r_0/L^2\).

To solve the wave equation in general, it is convenient to define the coordinate
\be v = 1 - \frac{r_0^2}{r^2} \ee
which goes from \(v = 0\) at the wall (\(r = r_0\)), to \(v = 1\) at the boundary (\(r\to\infty\)). The wave equation in terms of the new coordinate is
\be 4 v (v^3 - 4v^2 + 6v -3)\psi^{\prime\prime}(v) + 4(v^3-3v^2+3v-3)\psi^\prime(v) - \left(\mu^2 + \frac{432(1-v)^5}{(2v^3-6v^2+6v+3)^2}\right)\psi(v) = 0 \ee
with \(\mu\) as above.

In this model one cannot take Neumann or Dirichlet boundary conditions at the wall. If \(\psi\) vanishes at the wall then by the wave equation so does its first derivative, and vice versa. The only solution satisfying both those conditions is that \(\psi=0\) everywhere. Instead, the boundary conditions one imposes are only regularity at the wall and the vanishing of the wave function on the boundary. That is enough to get a discrete spectrum. 

\begin{figure}\centering
	\includegraphics[width=0.6\textwidth]{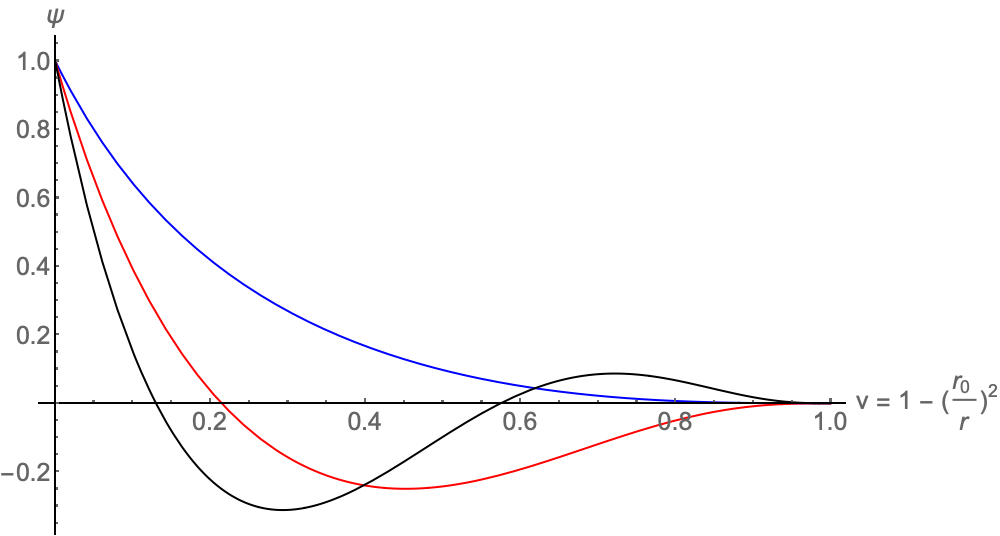}
	\caption{Wave functions of the first three scalar modes in Witten's model (here normalized such that \(\psi(0) = 1\)). \label{fig:psi_Witten}}
\end{figure}

It is possible to solve the wave equation formally in terms of a power series in \(v\), with a recursive definition for the coefficients. For the following, it is simpler to solve the equation for \(\psi(v)\) numerically using the shooting method, obtaining the spectrum first derived in \cite{Brower:2000rp}. The first few values are
\be \mu^2 = 7.308\,,\qquad 46.985\,,\qquad 94.482\,,\qquad 154.963\,,\qquad 228.709\,\ldots \ee
The spectra found in this model are not of Regge form. They are better approximated by a KK-like spectrum of the form \(m^2 \approx a n^2 + b n + c\) for some non-zero constants \(a\), \(b\), \(c\). The corresponding wave functions are drawn in figure \ref{fig:psi_Witten}. Note that they all have a maximum at \(r = r_0\).

It was shown that the spectrum of glueballs derived from the 11D background and from the 10D compactified $D4$ background are the same. It is not obvious that the scattering amplitude of closed strings in these two backgrounds will be the same. This question will be tested in the following.

\subsection{Generalized confining background} \label{sec:general_background}
One can study more general 10 dimensional backgrounds of the form
\be ds^2 = g_{4}(U) \eta_{\mu\nu}dx^\mu dx^\nu + g_U(U) dU^2 + g_{T}(U) d\Omega_5^2 \ee
where we have the four dimensional coordinates $x^\mu$, a radial coordinate \(U\), and five compact coordinates. For notational simplicity we take the compact space to be an \(S_5\), though that is not necessary and will not be the case for all examples considered in the following.

There are two simple conditions that the metric can satisfy that will ensure that the dual gauge theory be confining, in the sense of an area law for Wilson loops \cite{Kinar:1998vq}. The conditions are:
\begin{enumerate}
\item \(g_4^2\) has a non-zero minimum at \(U=U_\Lambda\)

\textbf{or}
\item \(g_4 g_U\to\infty\) at \(U_\Lambda\) and \(g_4(U_\Lambda)\neq0\).
\end{enumerate}
In hard wall models one imposes the first condition by hand, while Witten's model is an example of a confining background fulfilling the second condition.

For closed string scattering, we will be interested in the wave function of the lightest bulk scalar mode in such backgrounds. The wave function \(\psi(U)\) is then the radial dependence of the solution to the KG equation for a `massless' scalar propagating in the given confining background, where for simplicity one neglects any dependence on the compact coordinates:
\be \Box \Phi = 0\,, \qquad \Phi = \Phi(x^\mu,U) = e^{i \eta_{\mu\nu}k^\mu x^\nu} \psi(U) \ee

More specifically, we would like to determine the behavior far from the IR wall. For a qualitative analysis in that regime let us assume now that all the components of the metric scale as some power of \(U\),
\be g_4(U) \sim U^{\delta_4}\,, \qquad g_U(U) \sim U^{\delta_U}\,, \qquad g_T(U) \sim U^{\delta_T} \ee
Where \(\delta_4<0\) since the UV is at large \(U\). The solution at large \(U\) then is asymptotically given by a Bessel function, as in \(AdS\),
\be \psi_i(U) \to U^{-\frac{\Delta_i}2}J_{\frac{\Delta_i}{\eta}}(m U^{-\frac{\eta}2}) \ee
where the mass scale \(m\) is determined by the background, and the other parameters are
\be \Delta_i = \frac12(4\delta_4+5\delta_T - \delta_U) - 1\,, \qquad \eta = \delta_4 - \delta_5 - 2 \label{eq:general_background_delta}\ee
They are both taken to be positive, as is the case in \(AdS_5\) where \(\Delta =4\) and \(\eta = 2\). The wave function for the lowest scalar mode is expected to behave as \(U^{-\Delta}\) at large \(U\).

\subsubsection{Backgrounds with non-trivial dilaton}
In some holographic models of QCD the background fields include a non-trivial dilaton profile, in the sense that the dilaton becomes an active field that depends on the radial coordinate as in many RG flows see e.g. \cite{Bianchi:2001de, Bianchi:2001kw, Gursoy:2007cb, Gursoy:2007er, Gursoy:2008za}. In this case we need to incorporate it into the integration over the radial coordinate, generalizing eq. \eqref{eq1} as
\be {\cal A}_4(s,t,u) = \int_{U_\Lambda}^\infty dU \sqrt{-g} e^{-2\phi} \big(\prod_{i=1}^4 \psi_i(U)\big) {\cal A}_{10}(\tilde s,\tilde t,\tilde u) \label{eq:a_dilaton} \ee

\subsection{Introducing fundamental quarks} \label{sec:quarks}
In most of the following we will deal with closed strings, which correspond to glueballs. However, open strings corresponding to mesons and baryons are of more immediate phenomenological significance, so we will briefly describe the introduction of quarks into holographic QCD models.

Since the early days of string theory it has been understood that fundamental quarks should correspond to (the ends of) open strings. In modern terms, this calls for D-branes.
It is thus natural to wonder, whether one can consistently add 'flavor' D-brane probes to
supergravity backgrounds duals
of confining gauge theories, that can accommodate open strings and fundamental quarks.
In case that the number of flavor D-branes is small $N_f \ll N_c$   one can convincingly argue that their
back-reaction on the bulk geometry is negligible so that they can be treated in a probe approximation.
It was also well known that open strings between parallel $N_f$ D7-branes and $N_c$ D3-branes play the role of flavored quarks (hypermultiplets) in the $SU(N_c)$
gauge theory that resides on the D3-brane 4d world volume. Karch and Katz \cite{Karch:2002sh}
 proposed to elevate this brane configuration into a supergravity background by introducing probe
D7-branes  into the $AdS_5\times S^5$ background\footnote{A full back-reacted geometry would require the introduction of orientifolds $\Omega$-planes to avoid conical singularities and axio-dilaton monodromies at infinity \cite{Aharony:1999ti, Bianchi:2013gka, Addazi:2015rwa, Addazi:2015hka, Bianchi:2020fuk}. A `third' scenario may take place in this setting \cite{Antinucci:2020yki,Antinucci:2021edv,Amariti:2021lhk,Etxebarria:2021lmq}. }. Note that these types of  setups are  non-confining ones.

This idea was further explored in \cite{Kruczenski:2003be} and in many other followup works. For a list of them see the review \cite{Erdmenger:2007cm}. There have been certain attempts to go beyond the flavor probe approximation for instance \cite{Burrington:2004id}. The first attempt to incorporate flavor branes in a confining background was in \cite{Sakai:2003wu} where $D7$ and $\overline{D7}$-branes were introduced to the Klebanov-Strassler model \cite{Klebanov:2000hb}. This program was later completed in \cite{Dymarsky:2009cm}. An easier setting for incorporating flavor branes is the Witten-Sakai-Sugimoto (WSS) model \cite{Sakai:2004cn,Sakai:2005yt,Rebhan:2014rxa}. One starts with a pair $D8$-$\overline{D8}$ located at diametrically opposite points on the `thermal' circle $x_4$ on the UV boundary. The two smoothly join in the IR where the geometry is capped (cigar shape). This is the geometric realization of the chiral symmetry breaking \(U(N_f)\times U(N_f)\to U(N_f)\).

For reference, we note that the induced metric on D8 is given by
\be
ds^2_{D8}  = \frac{U^{3/2}}{L^{3/2} }\left(dx^2 + \left[f(U) + \frac{L^3}{U^3 f(U)} \left(\frac{dU}{dy}\right)^2\right]dx_4^2\right) + \frac{L^{3/2}}{U^{3/2}} U^2 d\Omega_4^2 
\ee
that accounts for the bending of the profile of the probe D8. Here \(U = U(x_4)\) is the profile of the brane.

Flavor was also introduced in the bottom-up models confining models. For instance the Veneziano limit of large $N_f$ in addition to the large $N_c$ was studied in \cite{Jarvinen:2011qe}.

%%%%%%%%%%%%%%%%%%%%%%%%%%%%%%%%%%%%%%%%%%%%%%%%%%%%
\section{Glueball scattering \`a la Polchinski-Strassler} \label{sec:Polchinski_Strassler}

In order to reproduce the hard `partonic' behavior of glueball scattering at large energies in a holography-inspired hard wall model, Polchinski and Strassler \cite{Polchinski:2001tt}, 
assumed that the scattering process be dominated by the interior region near the IR wall, far from the (conformal) boundary (UV) region. As a result they concluded that the full amplitude might be given by a closed-string amplitude in flat spacetime with rescaled momenta that is then integrated over the radial/holographic variable. 

Using as holographic coordinate $r=L^2/z$, so that the IR hard-wall is located at $r=r_0=L^2/z_0$, and denoting by $\tilde{p}$ the 10-dimensional momenta and by $p$ the `inertial' momenta one has 
\be
\sqrt{\alp } \tilde{p} = \frac{r_0}{r} \sqrt{\hat{\alpha}^\prime} {p} \quad {\rm and} \quad  r_0 = \Lambda L^2
\ee

Glueball scattering amplitudes would then be given by 
\be
{\cal A}_{QCD}(s,t) = \int_{r_0}^\infty  dr d^5\Omega \sqrt{|G|} \prod_{i=1}^4\psi_i(r,\Omega) A_{string}(\tilde{s},\tilde{t})
\ee
where
\be
\psi_i(r,\Omega) = f(r) {\cal Y}(\Omega) \quad {\rm with} \quad f(r) \sim {r^\Delta_0 \over r^\Delta} \quad {\rm for} \quad r\rightarrow\infty
\ee
${\cal Y}(\Omega)$ some (spherical) harmonics and 
\be
{\cal A}_{string}(\tilde{s},\tilde{t}) = K_8(\tilde{p},h){\Gamma(1- {\alp  \over 4} \tilde{s}) 
\Gamma(1- {\alp  \over 4} \tilde{t}) \Gamma(1- {\alp  \over 4} \tilde{u})\over \Gamma(1+ {\alp  \over 4} \tilde{s}) 
\Gamma(1+ {\alp  \over 4} \tilde{t}) \Gamma(1+{\alp  \over 4} \tilde{u})}
\ee
where
\be
K_8(\tilde{p},h) = G_N { {\cal R}^4\over stu}
\ee
with $ {\cal R}_{\mu\nu\rho\sigma} = k_{[\mu} h_{\nu][\rho} k_{\sigma]}$ the `linearized' Riemann tensor of the metric fluctuation.

Neglecting the internal angular part ${\cal Y}(\Omega)$, one gets
\be
{\cal A}_{QCD}(s,t) \approx {g^2 \alpha^{\prime 3} \over L^6 r_0^4} \int_{r_0}^\infty  dr r^3 \prod_{i=1}^4{r^{\Delta_i}_0 \over r^{\Delta_i}} A_{string}(\tilde{s},\tilde{t}) \approx 
{(gN)^{\Delta - 2\over 4} \over N^2} {\Lambda^{\Delta-4} \over p^{\Delta-4}} 
\ee
where $\Delta=\sum_i \Delta_i$. 

As originally discussed by Polchinski and Strassler in \cite{Polchinski:2001tt} and  later on confirmed by other authors, in order to reproduce the expected QCD result incorporating the `partonic' behavior at very high energy, the scaling dimensions $\Delta_i$ should be replaced by the `twists' $\tau_i = \Delta_i - s_i$ which, up to corrections in $g_s$, is equal to number of constituents of the hadron. The lowest value is $\tau = 2 + ...$ both for mesons and for tensor glueballs. Scalar glueballs (dual to the axio-dilaton for instance) are to be treated as `twist 2' operators (lowest twists) even if
$\tau_{naive} = 4 + ...$. This subtle issue has never been fully clarified but it seems to be the only phenomenologically consistent choice.  

In their original analysis Polchinski and Strassler focused only on the high energy behavior that should replace the soft UV stringy behavior in flat spacetime with the hard behavior in curved AdS-like backgrounds dual to confining theories. They did not consider the poles that any string amplitudes exposes due to the infinite tower of Regge resonances. We will address this issue later on in our investigation with the caveat that the analysis of small $s$ regimes might be questionable in a simplified hybrid approach \`a la  Polchinski-Strassler.

\section{Glueball scattering as closed string scattering} \label{sec:closed}
In this section we will study binary \(2\to 2\) scattering of glueballs in terms of closed strings in various holographic models. These models are the  bottom up models of (i) the hard wall, (ii) the  soft wall 
  and the top down model referred to as  Witten's model \cite{Witten:1998qj} which is the near horizon limit of  the background of  large number of M5 branes compactified on a circle and then dimensionally reduced or alternatively, that of   the large number of   compactified D4 branes \cite{Brandhuber:1998er}.
	As was discussed in the introduction, in such holographic models the duals of the gauge theory 
 glueballs, such as the scalar operator $\textrm{Tr}F^2$  can be taken to be either a `massive' closed string excitation in these backgrounds or as a fluctuation of the `massless' supergravity fields, like the dilaton $\Phi$, the metric  and  the RR field, or their KK excitations.  Following  \cite{Polchinski:2001tt}, we will employ a hybrid formalism where the scattering amplitude is derived from  that    in flat spacetime with rescaled momenta that is then integrated over the radial/holographic direction weighted by the product of the wave functions  associated with the bulk fields which are  dual to the glueballs.
 
Apart from studying several holographic models we also generalize the work of \cite{Polchinski:2001tt} by using three type of computations of the scattering amplitude: (i) by  expanding around the poles of the flat spacetime amplitude, (ii) taking  the fixed angle approximation, and (iii) determining the Regge behavior. The first type of computations enables us to study the fate of the bound states that are being exchanged in the scattering process.  
 %variable that is based on assume that the scattering process is dominated by the interior region near the IR, which is far from the boundary (UV) region. As in the previous section we make the assumption that the full amplitude is given by the string amplitude in flat spacetime with rescaled momenta that is then integrated over the radial/holographic variable. After confirming the presence of the hard `partonic' behavior at large energies, we will study the fate of the Regge poles as well as other issues such as factorization and optical theorem. 

\subsection{Calculating the amplitude and its limits}
For closed strings, in order to avoid cumbersome tensorial structure, one can consider the four dilaton amplitude in the ten dimensional (flat space) superstring theory. We use the Virasoro-Shapiro-Brink-Green-Schwarz amplitude \cite{Virasoro:1969me}, which has the form.\footnote{We will not keep track of most numerical factors, for notational simplicity.}
\be
{\cal A}_{10}(s, t,u)= \left((\alp  s)^{4} +(\alp  t)^{4}+(\alp  u)^{4}  \right) {\Gamma(-\alp  s/4)\Gamma(-\alp  t/4)\Gamma(-\alp  u/4) \over \Gamma(1+\alp  s/4)\Gamma(1+\alp  t/4)\Gamma(1+\alp  u/4)} \label{eq:VirasoroShapiro}
\ee

The four dimensional scattering amplitude, following the prescription of \cite{Polchinski:2001tt}, can be written as
\be
{\cal A}_4 (s,t,u) = \int_{r_0}^{\infty}  dr \sqrt{-g}  \psi(r)^4\,{\cal{A}}_{10}(
\alp\tilde s(r),\alp\tilde t(r),\alp\tilde u(r)) \label{eq:A4ansatz}
\ee
where $r$ denotes the holographic coordinate, $r_0$ the lower bound of the integral will be specified for each model and $g$ denotes the determinant of the holographic metric, which is also a function of \(r\). We use the same wave function \(\psi(r)\) for each of the scattered states.

The crucial feature here is the rescaling of momenta and hence the Mandelstam variables with the radial coordinate. The rescaling factors can be read directly from the metric,
\be\label{rescaling}
%p_{\mu}\rightarrow \tilde p_\mu = {R\over r}p_{\mu} \qquad \Rightarrow \qquad 
s\rightarrow \tilde s =   g^{xx}s
\ee where $x$ is the direction of the momentum and $g^{xx}$ is the inverse of the metric in that direction.
The same rescaling also occurring for \(\tilde t\) and \(\tilde u\). The Mandelstam variables are always multiplied by \(\alp\), so one can equivalently think of the factor of $g_{xx} = 1/g^{xx}$ as a rescaling of the effective string tension of the scattered strings. For the lowest lying state $0^{++}$, or the operator $\text{Tr} F^{2}$, we expect the wave function \(\psi(r)\) to be the one of the ground state scalar mode propagating in the bulk. As was mentioned above to perform the integration we use three different techniques, to explore different regions and limits of the amplitude.
 
\subsubsection{Expansion around the poles}

In  this method we write the amplitude as an expansion in \(s\)-channel poles, perform the integral, and then sum the result. For the $s$-channel poles, the massless pole, $s=0$, produces $128 \alp {u^{2}\over s}  = 128 \alp  {t^{2}\over s }$.
The rest of the poles can be analyzed using the kinematic parametrization,
\be
t=-{s\over 2}(1-\cos\theta)\,,\quad u=-{s\over 2}(1+\cos\theta) 
\ee
where $\theta$ is the scattering angle. Then the amplitude can be written as
\be
{\cal{A}}_{10}=-4 (\alp s)^{3}(1+c_{+}^{4} +c_{-}^{4} ) {\Gamma(1-\alp s/4)\Gamma(\alp s\,c_{+}/4)\Gamma(\alp s\,c_{-}/4) \over \Gamma(1+\alp  s/4)\Gamma(1-\alp s\,c_{+}/4)\Gamma(1-\alp s\,c_{-}/4)}
\ee
where we use the shorthand notation
\be c_\pm = \frac12(1\pm \cos\theta) \ee
The $s$-channel poles are at $\alp s=4(n+1)$ and we can write the expansion
\be\label{expansion}
\begin{split}
{\cal{A}}_{10}&=4^{4}\sum_{n=0}^{\infty} (-1)^{n} {(n{+}1)^{2}\over (n!)^{2}} {(1+c_{+}^{4} +c_{-}^{4} )\over \alp s/4 - (n+1)} {\Gamma\big( (n{+}1)c_{+} \big) \Gamma\big( (n{+}1)c_{-} \big) \over \Gamma\big( 1{-}(n{+}1)c_{+} \big) \Gamma\big( 1{-}(n{+}1)c_{-} \big)}\\
&\equiv\sum_{n=0}^{\infty} {{\cal R}_{n}(\theta)\over \alp s/4 -(n+1)}
\end{split}
\ee
The residue at each pole ${\cal R}_{n}(\theta)$ is a function of the scattering angle, a polynomial of \(\cos(\theta)\). This is important when one rescales the Mandelstam variables, because as \(s\) and \(t\) change their ratio remains fixed. Therefore the angle and residue of the pole do not depend on the radial coordinate in the integral. Then, it is relatively simple to perform the integral over each pole, and write an expression for the full amplitude as
\be {\cal A}_4(s,t,u) = \sum_{n=0}^\infty {\cal R}_n(\theta) \int_{r_0}^\infty dr \sqrt{-g}\frac{\psi(r)^4}{\alp \tilde s(r)/4 -(n+1)} \label{eq:A_Pol} \ee
We will see that typically the integral evaluates to a hypergeometric function.

\subsubsection{The fixed angle approximation}
We examine the high energy behavior by using the fixed angle limit \(s\to\infty\) with fixed \(s/t\) of the original flat space amplitude, \eqref{eq:VirasoroShapiro}.

To give a brief derivation we use Euler reflection formula and Stirling approximation,
\be \Gamma(1-z)\Gamma(z) = \frac{\pi}{\sin(\pi z)}\,, \qquad \Gamma(z) \approx \sqrt{\frac{2\pi}{z}}\left(\frac{z}{e}\right)^{z} \ee
The former holds for any \(z\), while the later is an approximation when \(|z|\gg 1\) and \(\Rea[z] >0 \).

We use the first formula to make the arguments of the \(\Gamma\) functions positive, and then use Stirling's formula where \(s\), \(-t\), and \(-u\) are all taken to be large. Then the Virasoro-Shapiro amplitude becomes (we omit all constant prefactors)
\be {\cal A}_{10}^{\text{FA}} \approx \frac{\sin(\frac{\pi \alp t}{4})\sin(\frac{\pi \alp u}{4})}{\sin(\frac{\pi \alp s}{4})}\frac{(\alp s)^4+(\alp t)^4+(\alp u)^4}{stu} (\alp s)^{-\alp s/2}(-\alp t)^{-\alp t/2}(-\alp u)^{-\alp u/2} \ee
A compact way to write it is
\be {\cal A}_{10}^{\text{FA}} \approx \frac{\sin(\frac{\pi \alp t}{4})\sin(\frac{\pi \alp u}{4})}{\sin(\frac{\pi \alp s}{4})}\frac{(\alp s)^4+(\alp t)^4+(\alp u)^4}{stu}  e^{-2\beta_{stu}} \ee
with
\be \beta_{stu} = \frac {\alp s}4\log(\frac {\alp s}4) + \frac {\alp t}4\log(\frac {\alp t}4) +  \frac {\alp u}4\log(\frac {\alp u}4) \ee
We can write the same as a function of \(s\) and the scattering angle,
\be {\cal A}_{10}^{\text{FA}} \approx \frac{\sin(\frac{\pi}{4} \alp s c_{+})\sin(\frac{\pi}{4} \alp s c_{-})}{\sin(\frac{\pi}{4} \alp s)}{\cdot}\frac{1+c_{+}^4+c_{-}^4}{c_{+}c_{-}} {\cdot} s e^{-2\beta_{stu}} \ee
\be \beta_{stu} = -\frac {\alp s}4 \big(i \pi + c_{+} \log c_{+} + c_{-} \log c_{-}\big) \ee

The prefactor ratio of sine functions is a rapidly varying function which contains the zeros and poles of the amplitude. The last part, \(s e^{-2\beta_{stu}}\), is the usually quoted result and gives in a sense the average behavior of the amplitude at large energies. Since the prescription is now to replace \(s\to s(r)\) and integrate over \(r\), the rapidly oscillating function of \(s\) could a priori play a role in determining the high energy behavior of the amplitude. We verified by explicit computations of the integrals that omitting and considering only the average behavior does not alter the  scaling with \(s\) of the amplitude at high energies after integration. More physically, one can give \(s\) a small imaginary part in order to avoid the poles (and zeros) when taking the limit, and then take the average behavior.

Then, the behavior of the fixed angle limit of the four dimensional amplitude can be determined from the simple integral
\be {\cal A}_{4}^{\text{FA}}(s,\theta) \approx \frac{1+c_{+}^4+c_{-}^4}{c_{+}c_{-}}\int_{r_0}^\infty dr \sqrt{-g} \psi(r)^4 \tilde s(r) e^{-2\tilde \beta_{stu}(r)} \label{eq:A_FA} \ee
Since \(\beta_{stu}\) is proportional to \(s\), it scales in the same way with \(r\).

\subsubsection{The Regge regime} \label{sec:Regge}
Similarly to the fixed angle calculation, one can take the Regge limit of the amplitude, taking \(s\to\infty\) while keeping \(t\) fixed. The string amplitude becomes, up to a constant
\be {\cal A}_{10}^{\mathrm{R}}(s,t) = \frac{\sin[\frac14\pi\alp (s+t)]}{\sin(\frac14\pi \alp s)} \frac{\Gamma(-\alp t/4)}{\Gamma(1+\alp t/4)} \left(\frac{\alp s}{4}\right)^{\frac{\alp t}{2} + 2} \ee
As before we can leave out the oscillating prefactor, and for small \(t\) we can use the simpler form
\be {\cal A}_{10}^{\mathrm{R}}(s,t) \simeq \frac{1}{\alp t} \left(\frac{\alp s}{4}\right)^{\frac{\alp t}{2} + 2} \ee
We then need to evaluate the integral
\be  {\cal A}_{4}^{\mathrm{R}}(s,t) = \int_{r_0}^\infty dr \sqrt{-g} \psi(r)^4  {\cal A}_{10}^{\mathrm{R}}(\tilde s, \tilde t)\ee
To this end we will use a saddle point approximation, writing the integral as
\be
{\cal A}_{4}^{\mathrm{R}}(s,t) = \int_{r_{0}}^{\infty} dr \, e^{F_{st}(r)}
\ee
where
\be
F_{st}(r)= \log(\sqrt{-g} \psi(r)^4) - \log[\alp \tilde t(r)] + \left(2+\frac{\alp \tilde t(r)}{2}\right)\log\frac{\alp \tilde s(r)}{4}
\ee
We solve the saddle point equation $F^\prime_{s,t}(r)=0$, and then find
\be {\cal A}_{4}^{\mathrm{R}}(s,t) \simeq {e^{F_{st}(r^{*})}\over \sqrt{F^{\prime\prime}_{st}(r^{*})}} \label{eq:A_Reg}
\ee
where \(r^*\) is the location of the saddle point.

\subsection{The hard wall model} \label{sec:HardWall}
In the hard wall model, as described in section 2.1, the background metric is that of ordinary \(AdS_5 \times S_5\), with $r=L^2/z$, {\it viz.}
\be ds^2 = \frac{r^2}{R^2}\eta_{\mu\nu}dx^\mu dx^\nu + \frac{R^2}{r^2}dr^2 \label{eq:AdS_r} + R^2 d\Omega_5^2 \ee
except that the coordinate \(r\) now runs from a finite IR cut-off \(r_0 > 0\) (the ``wall'') to the boundary at infinity.
The ``warp factor'' now takes the form, per eq. \eqref{rescaling},
\be
p_{\mu}\rightarrow \tilde p_\mu = {R\over r}p_{\mu} \qquad \Rightarrow \qquad s\rightarrow \tilde s =  {R^{2}\over r^{2}}s
\ee

It is instructive to start by taking the wave function to have a simple power dependence on \(r\), namely
\be \psi_i(r) \simeq \left({r_{0} \over r}\right)^{\Delta_{i}} \ee
With this choice the four dimensional amplitude becomes
\be
{\cal A}_4(s,t,u) = \frac{r_0^\Delta}{R^4}\int_{r_0}^{\infty}  dr\,  r^{3-\Delta}  \,{\cal A}_{10}\big(\frac{R^2}{r^2}\alp s,\frac{R^2}{r^2}\alp t, \frac{R^2}{r^2}\alp u\big)
\ee
where the factor of \(r^3\) comes from the background $AdS_5$ metric, and
\be \Delta \equiv \sum_{i=1}^4 \Delta_i \ee
Notice that as in \cite{Polchinski:2001tt} we expect the scaling dimension $\Delta_{i}$ to be replaced by the twist $\tau=\Delta-s$ where $s$ is the spin.

\subsubsection{Expansion around the poles}
Using the pole expansion of the ten dimensional amplitude given in \eqref{expansion}, the 4-dimensional amplitude in the hard wall model becomes
 \be
{\cal A}_4(s,t,u) =4\frac{r_0^\Delta}{R^4}\sum_{n=0}^{\infty} {\cal R}_{n}(\theta)\, \int_{r_0}^{\infty}  dr\,    {r^{3-\Delta}  \over \frac{R^2}{r^2}\alp s -4(n+1)} \label{eq:HW_Pol}
\ee
The result of each of the integrals is a hypergeometric function
\be
\int_{r_0}^{\infty}  dr\,    {r^{5-\Delta}  \over R^2\alp s/4 -r^{2}(n+1)}=
\frac{r_0^{4-\Delta}}{(n+1)(4-\Delta)}{}_{2}F_{1}\Big(1\,; {\Delta\over 2}-2\,;{\Delta\over 2}-1\,; \frac{R^2}{r_0^2}{\alp  s\over 4 (n{+}1)}  \Big) \label{eq:2F1integral}
\ee
and we can write the full amplitude compactly as
\be
{\cal A}_4(s,t,u) =\frac{r_0^4}{R^4}\frac{1}{4-\Delta}\sum_{n=0}^{\infty} {{\cal R}_{n}(\theta)\over n{+}1}\, \,_{2}F_{1}\Big(1\,; {\Delta\over 2}-2\,;{\Delta\over 2}-1\,; {\alpe  s\over 4 (n{+}1)}  \Big) \label{eq:A_2F1}
\ee
We defined
\be \alpe = \frac{R^2}{r_0^2}\alp \ee
which is the effective value of the slope \(\alp\) in the dual four dimensional confining theory. In all models considered below the bare slope will be similarly replaced by an effective one depending on the scale set by \(r_0\).

Note that the integral in \eqref{eq:2F1integral} converges for \(\Rea[s] \neq 0\) or \(\alpe s < 4(n+1)\), that is, precisely away from the interesting physical region of real and positive \(s\). The hypergeometric function can be analytically continued to include the real positive axis as well.\footnote{Alternatively, one can shift the original poles by a small imaginary part by replacing \(n + 1 \to n + 1 + i\epsilon\) in eq. \eqref{eq:2F1integral} and then the integral will converge for any \(s\) on the real axis. We will do this in later sections.} For a given \(n\), there is a singularity at \(\alpe s = 4(n+1)\) but instead of a pole it is now a logarithmic branch-point. Therefore there is a branch cut starting at \(\alpe s = 4(n+1)\) and the amplitude will have an imaginary part for \(\alpe s > 4(n+1)\).

For even \(\Delta=2(k+1) > 2\), the hypergeometric function can be written explicitly as
\be {}_2 F_1(1\,;k\,;k+1\,;z) = -\frac{k \log(1-z)}{z^k} - \sum_{l=1}^{k-1}\frac{k}{(k-l)z^l} \label{eq:2F1k}\ee
which holds for any integer \(k\).\footnote{For general \(k\), the function can also be written in terms of the Lerch transcendent as \( {}_2 F_1(1\,;k\,;k+1\,;z) = k \Phi(z,1,k)\).}

Let us look more closely at the resulting amplitude \eqref{eq:A_2F1}, for even \(\Delta\). Substituting \eqref{eq:2F1k} yields
\be
{\cal A}_4(s,t,u) = \sum_{n=0}^\infty \frac{{\cal R}_n(\theta)}{n+1}\bigg(\log\big(1-\frac{\alpe s}{4(n+1)}\big)\frac{4^k k (n+1)^k}{(\alpe s)^k} +\sum_{l=1}^{k-1}\frac{4^l k(n+1)^l}{(k-l)(\alpe s)^l}\bigg) \label{eq:A_pol_explicit}
\ee
where \(k = \frac{\Delta}{2}-2\). 

The imaginary part of the amplitude comes from a single term, the logarithm which gives an \(i\pi\) for \(\alpe s > 4(n+1)\). Up to constants, it is
\be \Imm[{\cal A}_4] = s^{2-\frac{\Delta}{2}} \sum_{n=0}^{N_M} (n+1)^{\frac\Delta2-3} {\cal R}_n(\theta) \ee
where \(N_{M} = \lfloor\alpe s/4\rfloor\). When \(s\) is large, we can take as a good approximation \(N_M\to\infty\), then the infinite series can be written as
\be {\cal C}_l(\theta) \equiv \sum_{n=0}^\infty (n+1)^{l-1} {\cal R}_n(\theta) = -\frac{256 (1+c_{+}^{4} +c_{-}^{4})}{\pi^2} \sum_{m=1}^\infty \frac{m^{l+3}}{(m!)^2} \sin^2(\pi m c_{+}) \big(\Gamma(m c_{+})\Gamma(m c_{-})\big)^2 \label{eq:445}\ee
where we used identities of the Gamma function to bring \({\cal R}_n(\theta)\) to a more compact form. The above is a convergent series for any \(l\) and finite \(\theta\). It is divergent for \(\theta = 0\) or \(\theta=\pi\), which corresponds to the presence of the poles at \(t=0\) and \(u=0\).

As for the real part, one should be more careful when evaluating its high energy behavior, because of the summation over \(n\). For a finite number of poles one can easily see that the amplitude in \eqref{eq:A_pol_explicit} has the leading order term of \(s^{-1}\) at very high energy. In string amplitudes, the fact that there is an infinite number of poles plays a crucial role in the high energy behavior. Since there is an explicit \(1/s\) term in \eqref{eq:A_pol_explicit} and its coefficient does not appear to vanish in the summation, we argue that the behavior of the amplitude is
\be \Rea[{\cal A}_4] \sim s^{-1}\,, \qquad \Imm[{\cal A}_4] \sim s^{2-\Delta/2} \ee

For a unitary amplitude we expect that \(|{\cal A}|^2 \sim \Imm[{\cal A}](\theta=0)\). Assuming that the dependence of the imaginary part on $s$ holds for any value of $\theta$ and in particular also for forward scattering singles out the value \(\Delta = 8\), or \(\Delta_i = 2\) for each of the wave functions, giving $\Imm[{\cal A}] \sim s^{-2}$. The value \(\Delta_i = 2\) is then consistent with the naive interpretation of a bound state of two partons of $0^{++}\sim \mathrm{Tr} F^{2}$ . For general \(\Delta\) the amplitude does not seem to satisfy tree-level unitarity.

As we already mentioned $\Delta_i =2$ should be associated to the `twist', which is equal to the number of partonic constituents of the hadron rather than the dimension. Notice that in a realistic scenario for QCD there will also be an anomalous dimension that would shift slightly (in the UV) the value of $\Delta_i$ to $\Delta_i =2+\gamma_i$.

\begin{figure}[ht!] \centering
	\includegraphics[width=0.66\textwidth]{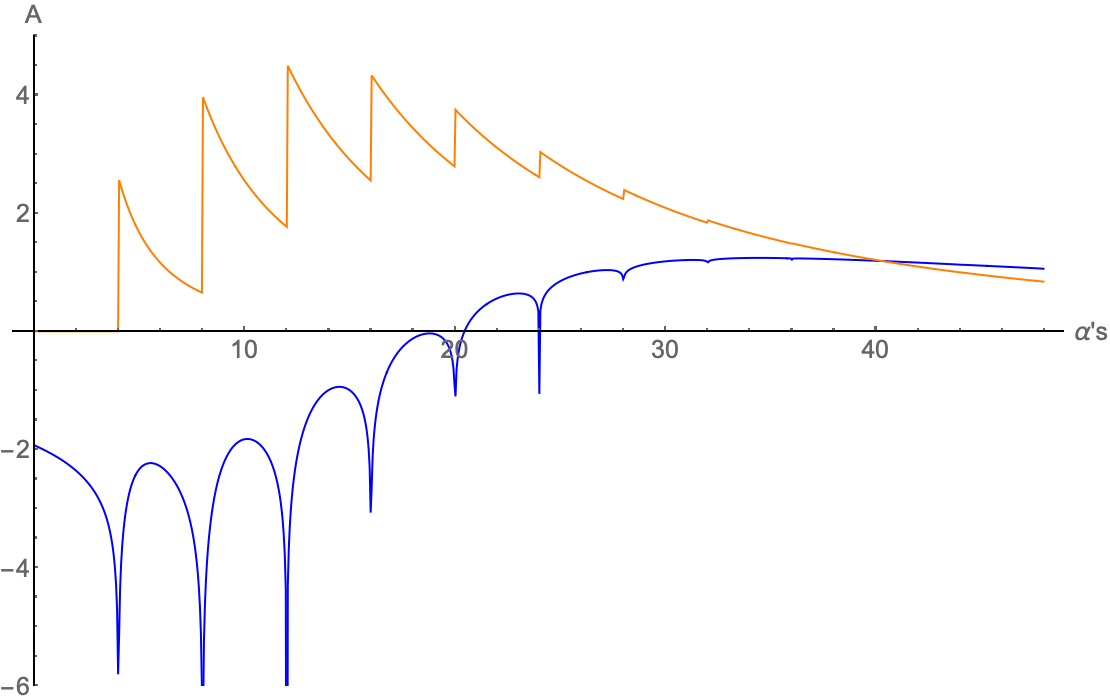}
	\caption{\textbf{Top:} Real (blue line) and imaginary (orange) parts of the amplitude computed in the hard wall model, eq. \eqref{eq:A_2F1}, with \(\Delta = 8\), and \(\cos\theta=0.8\). The plot shows a structure of singularities that fade away as we increase the energy.\label{fig:A_HardWall_Power}}
\end{figure}

One can also see from eq. \eqref{eq:2F1k} that the poles in the original amplitude are replaced with logarithmic singularities which are located at \(\alpe s = 4(n+1)\), as a consequence of the integration over the radial coordinate. The amplitude no longer has simple poles (except the one at \(s=0\)). We plot the resulting amplitude for a given angle and \(\Delta = 8\) in figure \ref{fig:A_HardWall_Power}.

Taking the integral \eqref{eq:HWFA} and inserting the wave function of \eqref{eq:psi_HW}, we compute numerically the integral
\be \int_0^1 dz \frac{\psi_0^4(z)}{z^5} \alp s z^2 e^{-2\beta_{stu} z^2} \ee
with \(\psi_0(z) = z^2 J_2(z)\), and find that it goes like \(s^{-6}\) for large \(s\). It is not surprising since for large \(s\) the integral is dominated by small \(z\) where \(\psi_0(z) \sim z^4\), and the result is in agreement with the previous result with \(\Delta = 4\times 4 = 16\). The issue is that the wave function uniquely determines that \(\Delta_i = 4\), which is the dimension, rather than the twist \(\Delta_i = 2\).

\subsubsection{Fixed angle regime}
Using eq. \eqref{eq:A_FA}, the fixed angle amplitude in the hard wall model is
\be
{\cal A}_4^{\text{FA}} \simeq \int_{1}^{\infty}dr\, r^{3-\Delta} \frac{\alp s}{r^2} e^{-2\beta_{stu}/r^{2}}={\alp s\over 2}\,\beta_{stu}^{1{-}{\Delta\over 2}} \left[\Gamma\left({\Delta\over 2}{-}1 \right)-\widetilde{\Gamma}\left({\Delta\over 2}{-}1; \beta_{stu}\right) \right]
\label{eq:HWFA}\ee
where $\widetilde{\Gamma}$ is the incomplete Gamma function. This latter term is exponentially suppressed at large \(s\), and the high energy behavior is determined to be
\be
{\cal A}_4^{\text{FA}} \simeq \int_{1}^{\infty}dr\, r^{3-\Delta} \frac{\alp s}{r^2} e^{-2\beta_{stu}/r^{2}} \simeq \Gamma\left({\Delta\over 2}{-}1\right){\alp s\over 2}\,\beta_{stu}^{1{-}{\Delta\over 2}}  \sim s \beta_{stu}^{1-\frac\Delta2} \sim s^{2-\frac\Delta2} \ee
In the computation based on the pole expansion this was the result for the imaginary part of the amplitude, while here we find that the full amplitude scales as \(s^{2-\Delta/2}\).

%%%%%%%%%%%%%%%%%%%%%%%%%%%%%%%%%%

\subsubsection{Regge regime}
As discussed in section \ref{sec:Regge}, we evaluate the integral
\be  {\cal A}_{4}^{\mathrm{Regge}}(s,t) = \int_{r_0}^\infty dr r^{3-\Delta}  {\cal A}_{10}^{\mathrm{Regge}}(\tilde s, \tilde t)\ee
by use of a saddle point approximation. The saddle point (at leading order at large \(s\)) is located at
%\be r^* \approx R\sqrt{\frac{\alp |t|}{\Delta-1}\log\left(\frac14 e(\Delta-1)\frac{s}{|t|}\right)} \ee
\be r^* \approx R\sqrt{\frac{\alp |t|}{\Delta-1}\log\left(\frac{s}{|t|}\right)} \ee
Then, the behavior in the Regge limit of the four dimensional amplitude is, at leading order,
\be {\cal A}_{4}^{\mathrm{Regge}}(s,t) \sim (\alp s)^2 (\alp |t|)^{-\frac\Delta 2}(\log \frac{s}{|t|})^{1-\frac\Delta2} \ee
which is the same behavior as the one found in \cite{Polchinski:2001tt}.

\subsubsection{Effect of exact wave function and boundary conditions} \label{sec:Bessel}
In the previous subsections we used a power \(r^{\Delta_i}\) in place of the wave function. In the hard wall model, the wave function has the simple form given in \eqref{eq:psi_HW}. For this wave function, we can compute the amplitude at a fixed angle \(\theta\) numerically as
 \be
{\cal A}^{HW}_{4pt}(s,t,u) =\sum_{n=0}^{\infty} {\cal R}_{n}(\theta)\, \int_{0}^{z_0}  dz\,   \frac{z^{-5} \psi_0(z)^4} {z^2 \alp s -4(n+1+i\epsilon)} \label{eq:a_hw_exact}
\ee
where \(\psi_0(z)\) is the wave function of the ground state. The expansion in \(s\)-channel poles is not strictly necessary for a numerical computation, but it improves the stability of the numerical integration algorithms. To avoid the singularities in the numerical integration we add a small imaginary part via the replacement \(n \to n+i\epsilon\) in the denominator, physically giving a small width to each resonance.%\footnote{In the previous subsection, where we used \(z^\Delta_i\) in place of the wave function, the integral of eq. [4.9] formally converges to the hypergeometric function on the RHS only for \(\alp s < 4(n+1)\) or away from the real axis. There we analytically continued the result by using the same \({}_2 F_1\) function for any \(s\), which gives the same result as an \(i\epsilon\) prescription.}

\begin{figure}[ht!] \centering
	\includegraphics[width=0.66\textwidth]{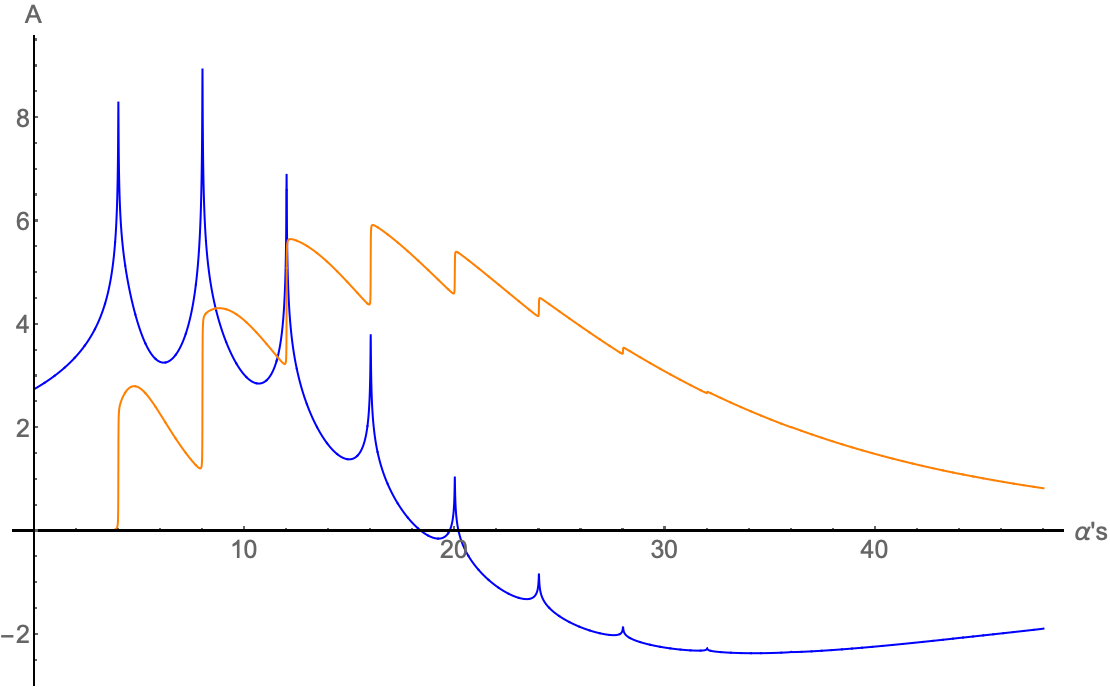} \\
	\caption{\textbf{Top:} Real (blue line) and imaginary (orange) parts of the amplitude computed in the hard wall model. The parameters used are \(\cos\theta = 0.8\), \(\kappa = 3.832\), and \(\epsilon=0.005\).\label{fig:A_HardWall}}
\end{figure}

The result for the amplitude is plotted in figure \ref{fig:A_HardWall}. We observe the same structure of low-energy logarithmic singularities, slightly smoothed by the \(i\epsilon\) regularization. The behavior at high energies observed in the numerical calculation is that \(\Rea[{\cal A}] \sim s^{-1}\) and \(\Imm[{\cal A}] \sim s^{-6}\). The behavior as \(s^{-6}\) is expected because the wave function \eqref{eq:psi_HW} behaves as \(z^4 \sim r^{-4}\) far from the wall. The dimension of the operator dual to a massless scalar in \(AdS_5\) is known to be \(\Delta_i = 4\), for the total \(\Delta = 16\). The choice of wave function is then in contradiction with the notion of taking \(\Delta\) to be the twist, rather than the dimension, and leads to a softer behavior that we would like. The results in the fixed angle and Regge limits are as in the previous subsection, if setting \(\Delta=16\).

\begin{figure}[h!]
\includegraphics[scale=0.48]{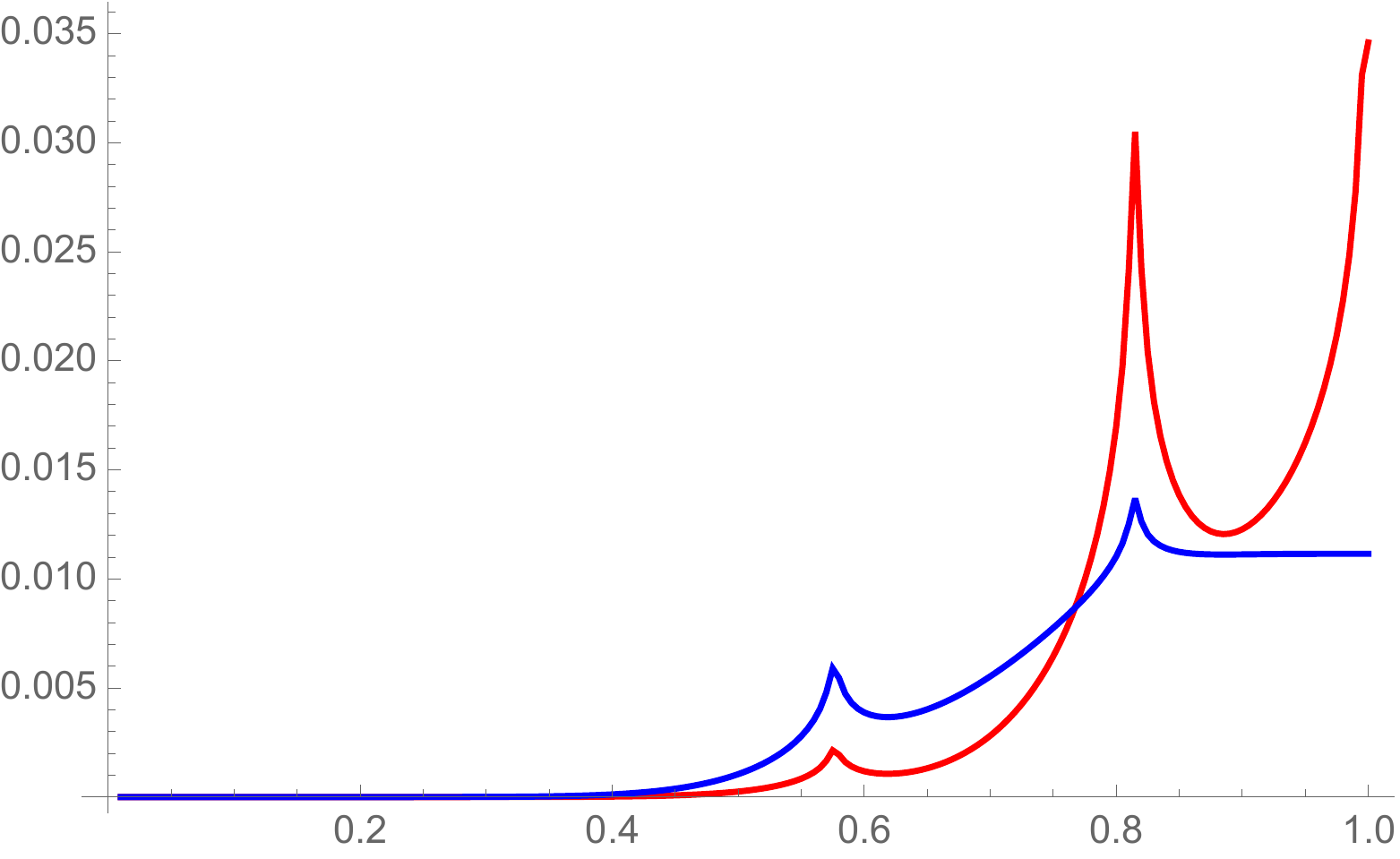}
\includegraphics[scale=0.48]{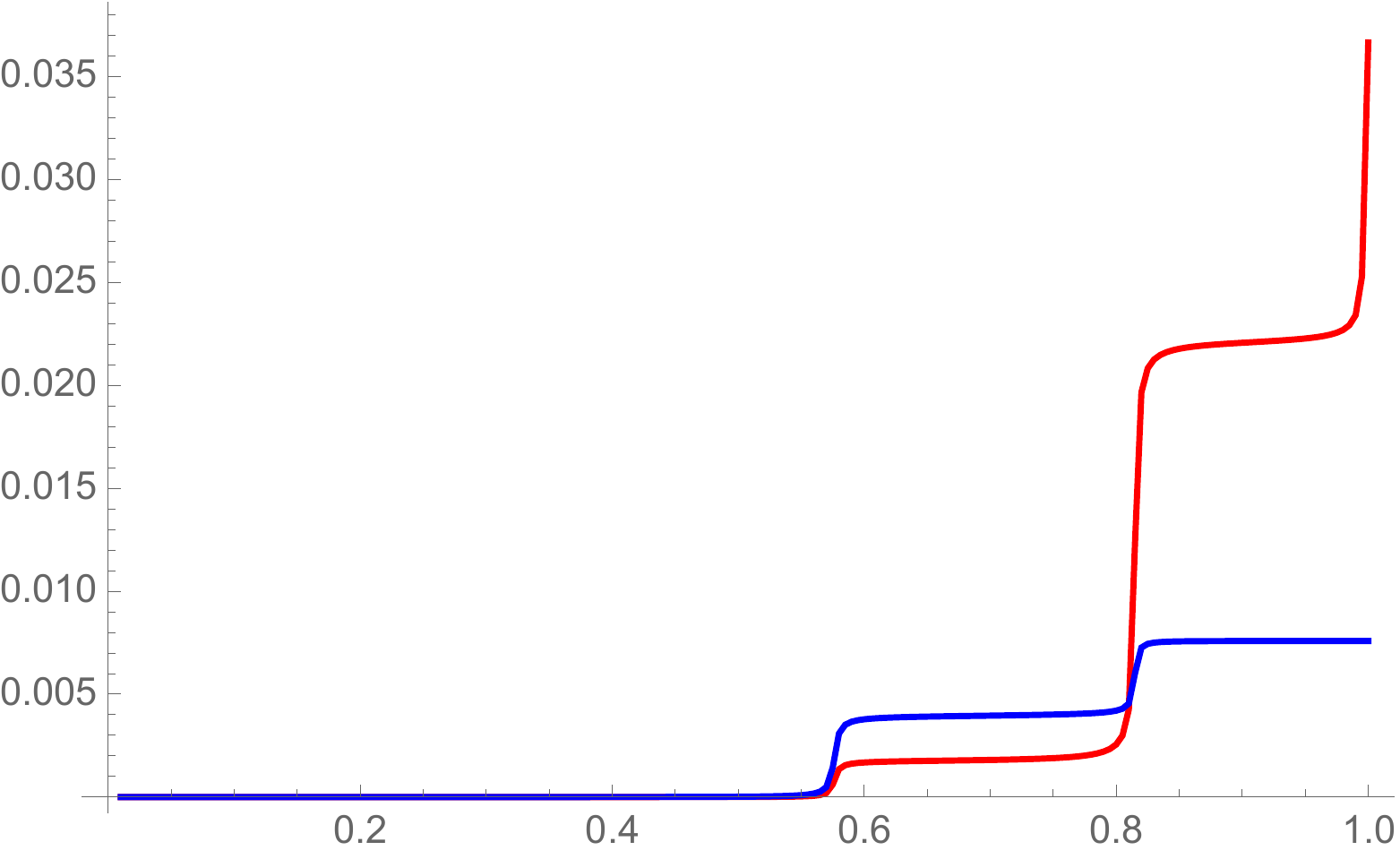}
\caption{\label{fig:dirichlet_vs_neumann} Real (left) and imaginary (right) parts of the integral \(\int_0^{z^\prime}\) of the amplitude as a function of \(z^\prime\). Red is a taking the wave function with Neumann boundary condition (maximum at \(z=1\)), blue is Dirichlet, vanishing at \(z=1\).  The value of \(\alpe s\) is 12.05, near a pole.  Contribution from new poles is zero in the Dirichlet case because of the vanishing of the wave function at the wall. Contribution from poles that we have previously crossed is mostly imaginary (branch cuts).}
\end{figure}

In this calculation we can see that while the high energy behavior is the same, one can eliminate the low energy structure of logarithmic singularities from the amplitude by imposing Dirichlet boundary conditions at \(r=r_0\). This is because the would be singularity is multiplied by \(\psi(r_0) = 0\) upon integration. In figure \ref{fig:dirichlet_vs_neumann}, we plot the integral \eqref{eq:a_hw_exact} from 0 to \(z^\prime\) as a function of \(z^\prime\), when \(\alpe s\) is fixed to be just near a pole. For a given \(s\) there is a finite number of singular points in the domain of integration given by 
\be z_n^2 = 4(n+1)/\alp s\ee
As we increase \(s\), new singular points are crossed and they contribute as long as the wave function is not small near the wall. We see that for the Neumann case there is a large contribution from the new singularity near the wall at \(z = 1\), but this contribution is suppressed by the wave function in the Dirichlet case.

However, if the wave function does not vanish on the wall, such as in the hard wall model with Neumann boundary conditions or in Witten's model which we examine in section \ref{sec:Witten}, we do find the same structure of logarithmic singularities as in eq. \eqref{eq:A_2F1}, and that is a generic consequence of the integration over the radial coordinate.

%%%%%%%%%%%%%%%%%%%%%%%%%%%%%%%%%%

\subsection{The soft wall model}
In the soft wall model,  described in section \ref{sec:softwall}, the IR cutoff is represented by a ``dilaton'' background field. As a consequence there is a softening factor $e^{-\lambda^2 z^2}$ multiplying the square root of the  determinant of the metric and the  integral is taken over the whole range of the coordinate $z$.

\subsubsection{Expansion around the poles}
Upon expanding around the poles the amplitude reads
\be
 {\cal A}_{SW}(s,t,u)=\sum_{n=0}^{\infty} {\cal R}_{n}(\theta)\int_{0}^{\infty}  dz\,  e^{-\lambda^{2}z^{2}}  {z^{-5} \,\psi^{4}_{SW}(z)  \over   z^{2}\alp s - (n+1)}
 \ee
The wave functions in SW model are given by eq. \eqref{eq:psi_SW}
\be
\psi^{(i)}_{m_i,l}(z)=z^{\ell + 2} L_{m_i,\ell}(\lambda^2z^2)
\ee
For $m_i=0$ the Laguerre polynomial is $L_{0,\ell}(\lambda^2z^2)=1$, and in general, when we scatter four states of the degrees \(\{m_i\} = (m_1,m_2,m_3,m_4)\) one has the following integral
\be
 {\cal A}_{SW}(s,t,u)=\sum_{n=0}^{\infty} {\cal R}_{n}(\theta) \,W_{L} \int_{0}^{\infty}  dz\,  e^{-\lambda^{2}z^{2}}  {z^{-5+\Delta}  \over   z^{2}\alp s - (n+1)} \label{eq:A_SW_pol}
\ee
where the coefficients \(W_L\) and \(\Delta\) are given by
 \be
 W_{L}(\{m_{i}\})=\prod_{i=1}^{4} \sum_{j_{i}=0}^{m_i} {(-1)^{j_{i}}\over j_{i}!} \begin{pmatrix} m_i+\ell \\m_i-j_{i} \end{pmatrix}\,, \qquad
 \Delta(\{m_{i}\}) =8+4\ell+\sum_{i=1}^{4}m_{i}
 \ee
After the integration, the amplitude becomes
 \be
 \begin{split}
 {\cal A}_{SW}(s,t,u)=&-\sum_{n=0}^{\infty} {{\cal R}_{n}(\theta)\over2 (n{+}1)} \,W_{L}\exp\left(-{\lambda^{2}(n{+}1)\over \alp s}\right)   \left(-{\alp s \over n{+}1}\right)^{2{-}{\Delta\over 2}} \\
 %%%
 &\Gamma\left({\Delta\over 2}{-}2\right)\widetilde\Gamma\left(3{-}{\Delta\over 2}\,;-{\lambda^{2}(n{+}1)\over \alp s}\right)
 \end{split} \label{eq:A_SW_pol2}
\ee 
As in the hard wall, the integral does not converge when \(s\) is on the positive real axis. One can slightly shift the poles by a small imaginary part or simply use the analytic continuation provided by the function on the RHS .

In particular $\Delta \geq 8$ and it follows that $3 -\Delta/2$ can be a negative integer or negative half integer. Note that the amplitude now can be written as a \(\lambda\)-dependent prefactor times a function of \(\alp s/\lambda^2\). By choosing \(\lambda\) we set the scale and the effective slope is \(\alp/\lambda^2\).
 
 In the case in which $3 -\Delta/2=-n$ with $n\geq 1$, that means $\Delta$ even, one has the following representation for the incomplete gamma function
 \be
 \widetilde{\Gamma}(-n,z)={e^{-z}\over n!} \sum_{k=0}^{n{-}1} (-1)^{k} (n{-}k{-}1)! \,z^{k{-}n} + {(-1)^{n}\over n!}\widetilde{\Gamma}(0,z)
 \ee
 where
\be
\widetilde{\Gamma}(0,z)=-\gamma_{E} - \log z - \sum_{k=1}^{\infty}{(-z)^{k}\over k (k!)}
\ee 
While the case in which $\Delta$ is odd, that means $3 -\Delta/2=-(n{+}1/2)$ with $n\geq 1$, one can use the following representation 
\be
\widetilde{\Gamma}(-n-1/2,z)={e^{-z}\over (n+1/2)!} \sum_{k=0}^{n} (-1)^{k} (n{-}k{-}1/2)! z^{k{-}n{-}1/2} + {\pi \,(-1)^{n{+}1} \over (n+1/2)! } \big(1- \mathrm{erf} \sqrt{z}\big)
\ee
where
\be
\mathrm{erf}(z)={2 \over \sqrt{\pi}} \int_{0}^{z} dt \,e^{-t^{2}}
\ee
The lowest states are obtained from $\Delta=8+4\ell + \sum_{i=1}^{4}m_{i}$ with $\ell=0$ and in particular the leading behavior at large $s$ is obtained from the contributions with $\{m_{i}\}=0$, all the other contributions coming from $\{m_{i}\}\ne 0$ are sub-dominant at large $s$.

\begin{figure}[ht!] \centering
	\includegraphics[width=0.66\textwidth]{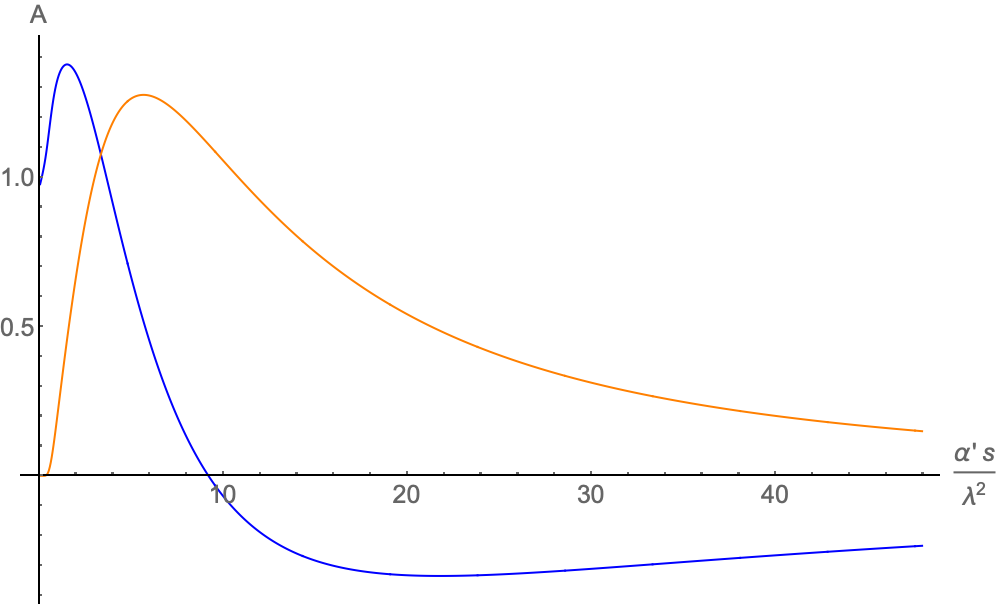} 
	\caption{\textbf{Top:} Real (blue line) and imaginary (orange) parts of the amplitude computed in the soft wall model. The parameters used are \(\cos\theta = 0.8\), \(\Delta = 8\). Both parts are smooth functions with no singularities.\label{fig:A_SoftWall}}
\end{figure}

The general result is
 \be
 \Rea[{\cal A}_{SW}]\sim s^{3{-}\Delta/2}\sim s^{-1} \quad \Imm[{\cal A}_{SW}]\sim s^{2{-}\Delta/2} \sim s^{-2}
 \ee
%%%%%%%%%%%%%%%%%%%%%%%%%%%%%%%%
The amplitude behavior for $\Delta=8$ is shown in figure \ref{fig:A_SoftWall}. We see that the result is a regular function with no apparent singularities. The disappearance of singularities here is mathematically more subtle than for the hard wall Dirichlet case explained in section \ref{sec:Bessel}. Since the integral in \eqref{eq:A_SW_pol} is from \(z=0\) to \(\infty\), it contains an infinite number of singularities for real, positive \(s\). When we shift the poles by \(i\epsilon\), or analytically continue the RHS of \eqref{eq:A_SW_pol2}, we find a smooth function.

\subsubsection{Fixed angle regime}
In the fixed angle regime one has to perform the integral
\be
\alp s\, \int_{0}^{\infty}dz\, e^{-\lambda^{2}z^{2}} z^{\Delta{-}5{+}2} e^{-2\beta_{stu}(z)}={\alp s\over 2}\,{\Gamma\left({\Delta\over 2}{-}1\right)} \left( \lambda^{2}+ 2\beta_{stu}\right)^{1{-}{\Delta\over 2}} \sim s^{2-\frac{\Delta}{2}}
\ee
The difference between the soft and hard wall models in this case is only in the shift of \(\beta_{stu}\) by \(\lambda^2\), which becomes irrelevant at scales \(\alp s\gg \lambda^2\).

\subsubsection{Regge regime}
In the Regge regime one needs to evaluate
\be
{\cal A}_{SW}^{\text{Regge}}=\int_{0}^{\infty}dz\, e^{-\lambda^2 z^2}z^{\Delta{-}5} {\cal A}_{\text{Regge}}(\tilde{s},\tilde{t})
\ee
The saddle point is at
\be
z^{*}\sim \left[ |t| \left(\log(s/|t|)-\frac{2\lambda^2}{t}\right) \right]^{-{1\over 2}}
\ee
Then the result is modified from the one in the hard wall model as
\be
{\cal A}_{SW}^{\text{Regge}}\simeq (\alp s)^{2}  (\alp |t|)^{-\Delta/2} \left(\frac{2\lambda^2}{|t|}+\log(s/|t|)\right)^{1-\Delta/2}
\ee
The \(s\to\infty\) limit is the same as in the hard wall, but the condition is that \(\log(s/|t|) \gg \lambda^2/\alp |t|\), or \(s\) is exponentially larger than \(|t|\).

%%%%%%%%%%%%%%%%%%%%%%%%%%%%%%%%%%
\subsection{Witten's model} \label{sec:Witten}
As explained in section \ref{sec:Witten_model} We have two different formulations of Witten's model. Starting from the 11-dimensional metric of \eqref{eq:AdS7}, the analysis of Witten's model is very similar to that in the hard wall model. The scaling of the Mandelstam variables is given by
\be \tilde s(r) = s r^{-2} \ee
as before, but owing to the fact that the background is asymptotically \(AdS_7\) we obtain a different behavior.

In the 10-dimensional formulation, in the background metric of \eqref{eq:backgroundWitten},
\be \tilde s(U) = s U^{-3/2} \ee
Since \(U\) is related to \(r\) by \(U = r^2\), the scaling is quite different. In the 10D version we also have to account for the presence of the non-trivial dilaton profile.

\subsubsection{11-dimensional model}
While the hard wall model and Witten's model are conceptually very different, the calculations of the amplitude are nearly identical. The biggest difference between this model and the \(AdS_5\) hard wall is due to the higher dimensional nature of Witten's model, namely that it is asymptotically \(AdS_7\times S_4\). The scaling of the Mandelstam variables is the same, but the integration measure now involves \(\sqrt{-g} \sim r^5\). In addition, the dimension of the lowest scalar mode is now \(\Delta_i = 6\) (see eqs. \eqref{eq:dimAdS} and \eqref{eq:dimWit}).

In the pole expansion method, taking a simple power dependence of the wave function for the time being,
 \be
{\cal A}_4(s,t,u) =\sum_{n=0}^{\infty} {\cal R}_{n}(\theta)\, \int_{r_0}^{\infty}  dr\,    {r^{5-\Delta}  \over r^{-2}\alp s -4(n+1)} \label{eq:A_Pol_Wit}
\ee
The only difference between this expression and the \(AdS_5\) version being an additional factor of \(r^2\) in the integral. The result is then
\be
{\cal A}_4(s,t,u) =\frac{1}{4-\Delta}\sum_{n=0}^{\infty} {{\cal R}_{n}(\theta)\over n{+}1}\, \,_{2}F_{1}\Big(1\,; {\Delta\over 2}-3\,;{\Delta\over 2}-2\,; {\alpe  s\over 4 (n{+}1)}  \Big) 
\ee
which for large $s$ behaves as
\be \Rea{{\cal A}}_4 \sim s^{-1}\,,\qquad \Imm{{\cal A}}_4 \sim s^{3-\frac\Delta2}
\ee
Using \(\Delta = 4\times 6 = 24\), one gets an imaginary part that behaves as \(s^{-9}\), a much higher power than in \(AdS_5\).

The fixed angle integral is
\be
{\cal A}_4^{\text{FA}} \simeq \int_{1}^{\infty}dr\, r^{5-\Delta} \frac{\alp s}{r^2} e^{-2\beta_{stu}/r^{2}}\approx {\alp s\over 2}\,\beta_{stu}^{2{-}{\Delta\over 2}} \Gamma({\Delta\over 2}{-}3)
\label{eq:A_FA_Wit
}
\ee
which again behaves as \(s^{3-\frac\Delta2}\).

The Regge limit is computed as
\be {\cal A}_{4}^R = \int_{r_0}^\infty dr r^{5-\Delta} {\cal A}_{10}^R \ee
The calculation is identical to the previous hard wall calculation if we shift \(\Delta\) by two. We use the same saddle point approximation and the result is that
\be {\cal A}_4^R \sim (\alp s)^2 (\alp |t|)^{1-\frac{\Delta}{2}} (\log \frac{s}{|t|})^{2-\frac\Delta2} \ee

To plot the amplitude at low energies, we use the exact wave function. This is done using the numerical solution of the wave equation \eqref{eq:psi_Wit}, where we now perform the integral in the coordinate \(v= 1-\frac{r_0^2}{r^2}\) as
 \be
{\cal A}_4 (s,t,u) =\sum_{n=0}^{\infty} {\cal R}_{n}(\theta)\, \int_{0}^{1}  dv\,   \frac{(1-v)^{-4} \psi_0(v)^4} {(1-v) \alpe s -4(n+1+i\epsilon)}
\ee
As in section \ref{sec:Bessel}, we shift the poles by \(i\epsilon\).

\begin{figure}[ht!] \centering
	\includegraphics[width=0.66\textwidth]{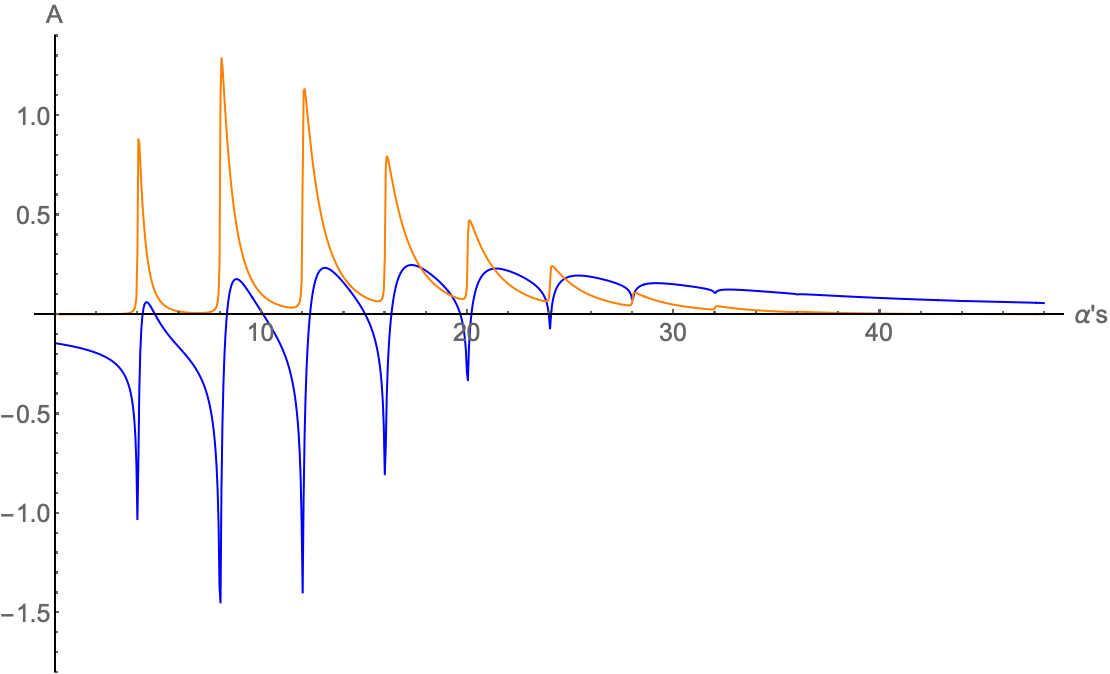}
	\caption{Real and imaginary parts of the amplitude computed in Witten's model. The parameters here are \(\cos\theta = 0.8\), \(\epsilon=0.005\).\label{fig:A_Witten}}
\end{figure}

The plots of the amplitude show once again the structure of low energy logarithmic singularities (figure \ref{fig:A_Witten}). In this model there are no Dirichlet boundary conditions, so we always have the singularities.

\subsubsection{10-dimensional model} \label{sec:Witten10}
The ten dimensional formulation of Witten's model yields different results. In terms of the variable \(U\), the momenta scale as \(\tilde s = U^{-3/2} s\).

In this model there is also the non-trivial dilaton, \(e^\phi \sim U^{3/4}\). This affects the scattering amplitude, as we now use the integration measure
\be dU \sqrt{-g}e^{-2\phi} = dU U^{4-\frac32} = dU U^{\frac52} \ee
in all integrals.

With these modifications we can repeat the calculations from the previous subsection. The integral over a single pole is
\be \int_{U_\Lambda}^\infty dU \frac{U^{5/2-\Delta}}{U^{-3/2} \alp s - 4(n+1)} = \frac{1}{4(n+1)(\Delta-\frac{11}{2})} {}_{2}F_{1}\Big(1\,; \frac23(\Delta-\frac{7}{2})\,;\frac23(\Delta-\frac{7}{2})+1\,; {\alpe  s\over 4 (n{+}1)}\big)  \ee
Summing over $n$ produces the behavior
\be \Rea[{\cal A}] \sim s^{-1}\,, \Imm[{\cal A}] \sim s^{\frac{7}{3}-\frac23 \Delta} \ee

As for \(\Delta\), the wave function of the scalar away from the wall behaves as \(U^{-3}\), for \(\Delta=12\).\footnote{This can be seen from the Klein-Gordon equation in this background at large \(U\), or from the wave equation for fluctuations of the dilaton around the classical solution. It is also compatible with the identification of \(U\) with \(r^2\) of the 11D model and the fact that \(\psi(r)\sim r^{-6}\) there.} Compared with the 11D calculation, the model now gives a conflicting prediction for the high energy behavior of the amplitude. With \(\Delta = 12\) we find that
\be \Imm[{\cal A}] \sim s^{-\frac{17}{3}} \ee
while in the 11D model the power was \(-9\).

The fixed angle result is
\be {\cal A}^{FA} \simeq \int_{U_\Lambda}^\infty dU U^{\frac52-\Delta}\frac{\alp s}{U^{3/2}}e^{-\beta_{stu} U^{-3/2}} \simeq \frac23 \Gamma\big(\frac23\Delta-\frac43\big) \, s \,\beta_{stu}^{-\frac23(\Delta-2)} \ee
which scales again as \(s^{-17/3}\) for \(\Delta=12\).

The Regge behavior is
\be {\cal A}^R = \int_{U_\Lambda}^\infty dU U^{\frac52-\Delta} {\cal A}^R(\tilde s,\tilde t,\tilde u) \sim (\alp s)^2 (\alp |t|)^{\frac13-\frac23\Delta} (\log \frac{s}{|t|})^{\frac43-\frac23\Delta} \ee

\subsection{A generalized background} \label{sec:general_background_closed}
We can repeat the calculations done above in a slightly more general background. Let us assume now the radial coordinate is called \(U\), with the momenta scaling as \(\tilde s = s U^{-a}\) and the wave function being \(\psi(U)^4 \sim U^{-\delta}\). The integral will also include \(\sqrt{-g}\) and possibly a dilaton factor \(e^{-2\phi}\) which when multiplied give \(U^b\).\footnote{By the assumption that \(e^{-2\phi}\) is a power of \(U\) we exclude the soft wall model discussed above, but include Witten's model.} In this case \(U\) is any coordinate such that the IR wall is at \(U=U_0 > 0\) and the boundary at infinity, such as \(r\) in \(AdS\). We  set \(U_0 = 1\), assume \(a>0\), and we also need \(\delta > 1+b-a\) to avoid UV divergences.

We can write the general expression for the integral we encounter when we perform the expansion in poles, it is
\be \int_{U_{\Lambda}}^\infty dU \frac{U^{b-\Delta}}{U^{-a}\alp s- 4(n+1)} = \frac{1}{4(n+1)(1+b-\Delta)} {}_{2}F_{1}\Big(1\,; {\Delta-b-1\over a}\,;{\Delta-b-1\over a}+1\,; {\alpe  s\over 4 (n{+}1)}\big)  \ee
which is a straightforward generalization of our previous results. This function now has a real part whose leading order behavior is \(s^{-1}\), and the imaginary part behaves as \(s^\alpha\), with
\be \alpha = \frac{1+b-\Delta}{a} \label{eq:alpha_ab}\ee

For any \(a>0\) we can always change to a coordinate \(r\) such that the scaling of Mandelstam variables is \(\tilde s = s/r^2\) as in \(AdS\), by the transformation \(r^2 = U^a\). The result of the integral does not depend on the frame since it is only a change of variables, and the physical scaling of the amplitude must remain the same under a coordinate change. Explicitly, In the new frame we get that \(a\to 2\), \(\Delta \to  \frac{2}{a}\Delta\), \(b \to \frac{2}{a}(b+1)\) and the power \(\alpha = (1+b-\Delta)/a\) remains the same.

If we want to associate \(\Delta\) with the sum of the dimensions of the operators in the dual theory, we could do it canonically in this ``\(AdS\) frame'' where \(a=2\), then the result is
\be \alpha = \frac{1+b}{2} - \frac{\Delta}{2} \ee

In \(AdS_5\), the remaining variable is \(b = 3\), and we get the by now familiar result \(\alpha = 2-\frac\Delta2\). This will also be the result for backgrounds that are asymptotically \(AdS_5\). For more general \(AdS_{d+1}\), we have that \(b = d-1\), and therefore
\be \alpha(AdS_{d+1}) = \frac d2-\frac\Delta2 \label{eq:alpha_adsd}\ee
In Witten's model (11D), which is asymptotically \(AdS_7\), we find this result with \(d=6\).  Only backgrounds that are asymptotically \(AdS_5\) give the desired behavior of \(s^{2-\Delta/2}\) for four dimensional QCD.

One can also repeat the analysis of the fixed angle and Regge regime integrals with the same type of background. For the fixed angle one has 
\be s\int_{U_{\Lambda}}^\infty dU U^{b-\Delta-a} e^{-\beta_{stu} U^{-a}} \approx  \frac{1}{a} s \beta ^{\frac{1+b-\Delta}{a}-1} \Gamma \left(1-\frac{1+b-\Delta}{a}\right) \sim s^{(1+b-\Delta)/a} \ee
So the general high energy dependence is as \(s^{\alpha}\) with the same power as in  eq. \eqref{eq:alpha_ab}, except that now it is not only for the imaginary part.

The result in the Regge limit is that
\be {\cal A}_4^R \sim (\alp s)^2 (\alp |t|)^{-2+\alpha} (\log \frac{s}{|t|})^{-1+\alpha} \ee
where \(\alpha = (1+b-\Delta)/a\) as before.

According to eq. \eqref{eq:general_background_delta}, one can express all the relevant quantities now in terms of the asymptotic behavior of the metric in the UV region. In the background of section \ref{sec:general_background} and assuming no dilaton, we should substitute \(\Delta = 4\Delta_i\), \(a = -\delta_4\), and \(b = \frac12(4\delta_4+\delta_U+5\delta_5) = \Delta_i + 1 + \delta_U\).

\section{Meson scattering as open string scattering } \label{sec:open}
Next we  would like to briefly describe the analysis of the scattering amplitudes  for open strings.   In holography, mesons are expected to be dual to open strings. This requires the introduction of flavor branes where open strings can end, corresponding to fundamental quarks in the dual theory. Usually the flavor branes are treated as probes of the holographic background created by the color branes.\footnote{This is not always the case, a notable exception is the ${\cal N} =2$ theory with $Sp(2N)$ gauge symmetry and $SO(8)$ global symmetry resulting from $N$ D3-branes at the unoriented singularity $C/Z_2$ with 4 D7-branes on top of an $\Omega 7^-$-plane.} Depending on the `embedding' of the flavor branes the fundamental quarks at the ends of the open strings can be massless or massive.

Lowest lying meson states correspond to quark bilinears such as the pseudo-scalar $\bar\psi_i \gamma_5 \psi^j$ or the vector currents $\bar\psi_i \gamma^\mu \psi^j$. 

We will focus on the massless pseudo-scalar case that leads to light-light mesons as pions that can be treated as (pseudo)Goldstone bosons so much so that they are expected to expose an Adler zero. The relevant description would be in terms of the so-called Lovelace-Shapiro amplitude \cite{Lovelace:1968kjy, Shapiro:1969km, Veneziano:2017cks}, that may be related to the non-supersymmetric fermion string \cite{Bianchi:2020cfc}.  We will also consider Veneziano (bosonic string) amplitude that was originally meant to describe the (decay) amplitude of the flavor singlet vector meson $\omega$ into three pions \cite{Veneziano:1968yb}.

The Polchinski-Strassler approach is more suited for closed string scattering, since it involves the wave-functions of bulk fluctuation modes. It is not immediately obvious how to account for the flavor branes and the structure of the open string. One can use the radial dependence of the eigenfunctions of the fluctuations of the branes, the modes one uses to calculate the meson spectra in holographic models.

However, if we take the wave function to behave as a simple power, \(\psi(r)_i \sim r^{\Delta_i}\), as we will do in the present work, then the analysis of the behavior of the holographic amplitude is very similar to the one found in the previous section for closed strings. The integrals over the radial coordinate have a nearly identical form. For this reason, we will mostly avoid repeating the calculations and simply cite results.

In this section, we will discuss three different amplitudes for open strings, the massless vector superstring amplitude \cite{Green:1982sw}, the Veneziano amplitude \cite{Veneziano:1968yb}, and the Lovelace-Shapiro amplitude \cite{Lovelace:1968kjy, Shapiro:1969km}. These all have the same high energy behavior, but different structures at lower energies.

\subsection{Superstring amplitude}
For simplicity, let's start by considering the open superstring, and in particular the 10 dimensional color ordered amplitude of four massless vector boson states,
\be
{\cal A}_{10}(s,t)=-{1\over 4 \alp{} ^{2} } {{\cal F}^{4}\over s t} {\Gamma(1-\alp  s) \Gamma(1-\alp t)\over \Gamma(1-\alp s -\alp  t)} \label{eq:A_open_vector}
\ee
where \cite{Bianchi:2015vsa}
\be
{\cal F}^{4}=2 \Tr{f_{1}f_{2}f_{3}f_{4}}- {1\over 2}\Tr{f_{1}f_{2}}\Tr{f_{3}f_{4}}+\text{cyclic in (234)}
\ee
with $f_{j}^{\mu\nu}=k_{j}^{[\mu}a_{j}^{\nu]}$ the usual field strength. 

In the holographic context massless vector bosons in the bulk correspond to conserved flavor currents $J^\mu$ in the boundary. In QCD these interpolate for vector mesons such as the $\rho$ meson. In a supersymmetric context these would have fermionic and (pseudo)scalar partners.  

The amplitude for 4-dimensional scalars $\phi_a$, with $a=1,... (d-4)$ is obtained choosing internal polarizations and 4-dimensional momenta $i.e$ $k_{j}{\cdot}a_{j}=0$ and $a_{j}^{M}\rightarrow \phi_{j}^{a}$. With these choices one has
\be
{\cal A}_{4p} =  \left({u\over s}(\phi_{1}{\cdot}\phi_{2})(\phi_{3}{\cdot}\phi_{4}) + (\phi_{1}{\cdot}\phi_{3})(\phi_{2}{\cdot}\phi_{4})+{u\over t}(\phi_{1}{\cdot}\phi_{4})(\phi_{2}{\cdot}\phi_{3}) \right){\Gamma(1-\alp  s) \Gamma(1-\alp t)\over \Gamma(1-\alp s -\alp  t)}
\ee
At large $s$ the dominant term of the expression is
\be
-{s\over t} (\phi_{1}{\cdot}\phi_{4})(\phi_{2}{\cdot}\phi_{3}) {\Gamma(1-\alp  s) \Gamma(1-\alp t)\over \Gamma(1-\alp s -\alp  t)}\Big|_{s{\gg}1} \label{eq:A_open_ss}
\ee

In the fixed angle regime the relevant part of the amplitude to take into account is given by
\be
{\cal A}_{4pt} \simeq -{s\over t}\, e^{-\beta_{st}}
\ee
with
\be
\beta_{st}=\alp s \log\alp s+\alp t \log\alp t - (\alp s+\alp t  ) \log(\alp s+\alp t )
\ee
As a function of the scattering angle $\theta$,
\be \beta_{st}=\alp s \Big(-c_{+}\log(-c_{+})- c_{-}\log (-c_{-})\Big)
\ee
which is essentially the same as in the closed string case, up to a rescaling of $\alp$ by a factor of 4.

\subsubsection{In the hard wall model}
Starting from the amplitude of eq. \eqref{eq:A_open_ss}, the expansion in poles in the \(s\)-channel is
 \be
 {\cal A}_{10}=\kappa(\theta) \sum_{n=0}^{\infty} {(-1)^{n}\over n!} {\Gamma\big(c_{+}(n{+}1)+1\big)\over \Gamma\big(c_{+}(n{+}1)-n\big)}{1\over \alp s-(n{+}1)}
 \ee
 where
 \be
 \kappa(\theta)= \left(-c_{-}(\phi_{1}{\cdot}\phi_{2})(\phi_{3}{\cdot}\phi_{4}) + (\phi_{1}{\cdot}\phi_{3})(\phi_{2}{\cdot}\phi_{4})+{c_{-}\over c_{+}}(\phi_{1}{\cdot}\phi_{4})(\phi_{2}{\cdot}\phi_{3}) \right) 
 \ee

In the hard wall background, the holographic amplitude is given by
 \be
 {\cal A}_4={1\over (4-\Delta)}\sum_{n=0}^{\infty} {(-1)^{n{+}1}\over (n+1)!} {\Gamma\big(c_{+}(n{+}1)+1\big)\over \Gamma\big(c_{+}(n{+}1)-n\big)}{\,_{2}F_{1}\Big(1\,; {\Delta\over 2}-2\,;{\Delta\over 2}-1\,; {\alpe  s\over n{+}1}  \Big)}
 \ee
The integral over a single pole gives the same result we had for the closed string. In the summation, the coefficient function (the residue) \({\cal R}_n(\theta)\) is different, and the poles are at \(\alp s = (n+1)\), rather than \(4(n+1)\). The behavior at high energies remains the same, 
\be \Rea[{\cal A}] \sim s^{-1}\,, \qquad \Imm[{\cal A}] \sim s^{2-\frac\Delta2} \ee

\begin{figure}[h!]
 \centering
\includegraphics[scale=0.5]{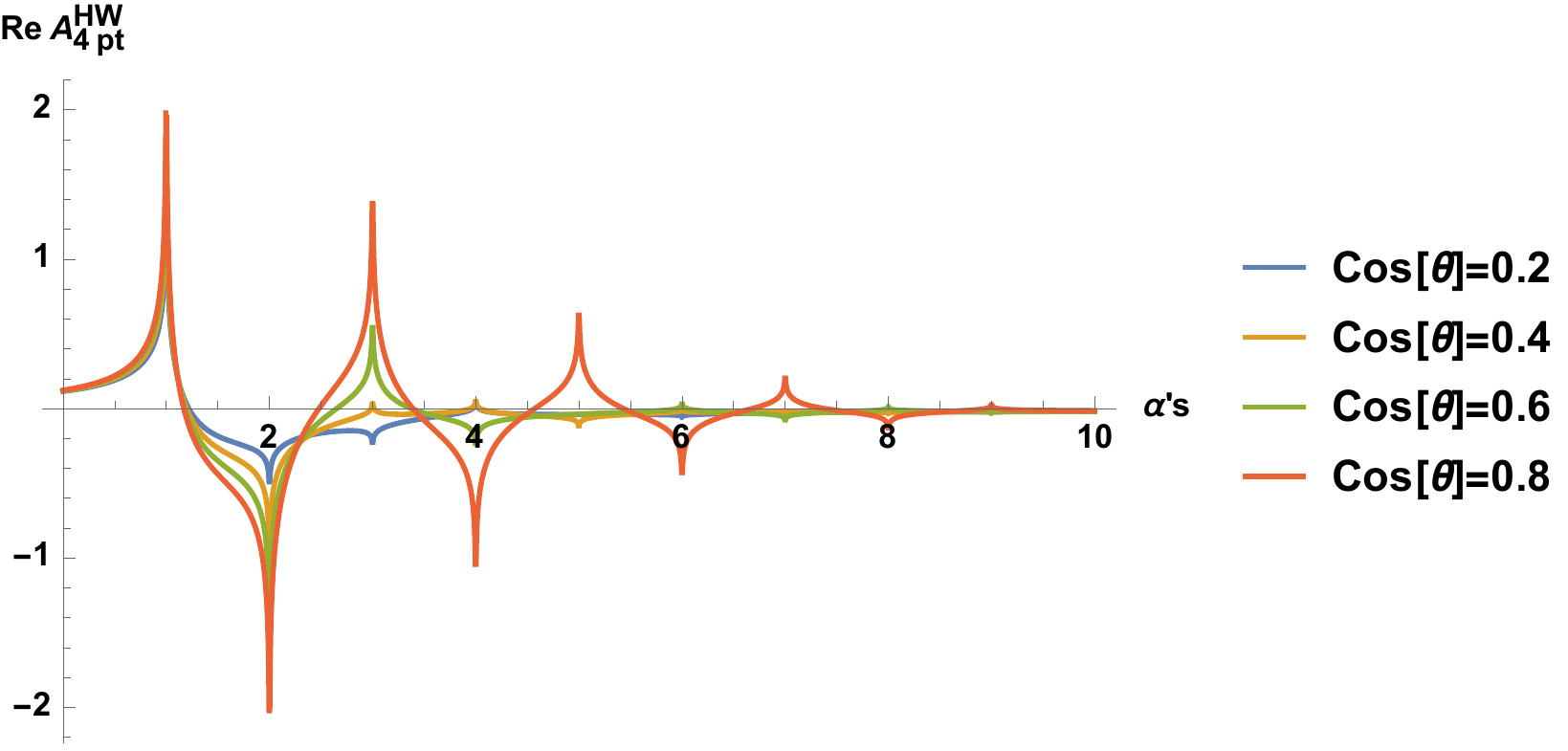}  
\includegraphics[scale=0.5]{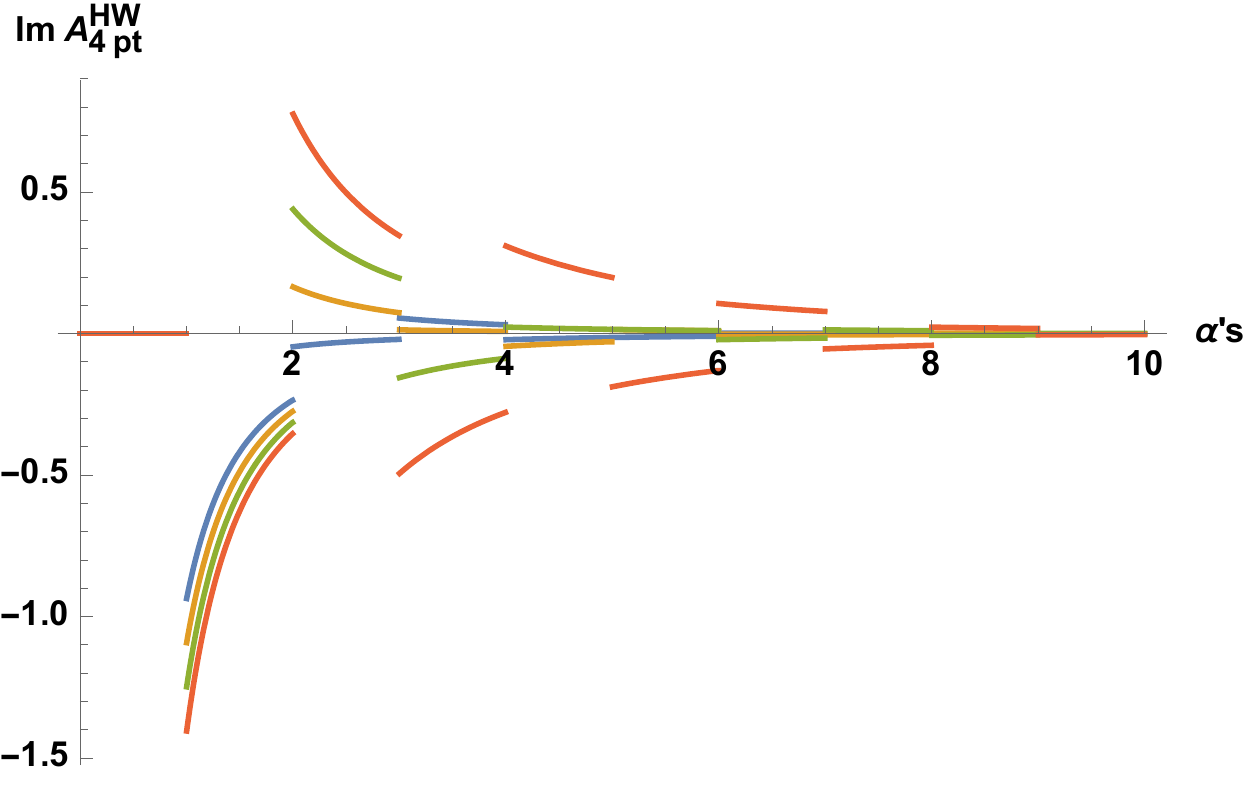}
 \caption{\label{fig:open_super_hw} Real and imaginary part of the open superstring amplitude in the hard wall background at different values of the angle.}
 \end{figure}

The amplitude for \(\Delta=8\) is drawn in figure \ref{fig:open_super_hw}.

In the fixed angle regime, one has
\be
{\cal A}_{4}^{\text{FA}} \simeq -{s\over t} \int_{1}^{\infty} dr\,r^{3-\Delta} e^{-{\beta\over r^{2}}}={1 \over 2 c_{+}} \beta_{st}^{2-{\Delta\over 2}} \left[\Gamma\left({\Delta\over 2}-2\right)-\widetilde{\Gamma}\left({\Delta\over 2}-2; \beta_{st}\right) \right]
\ee
Here the integral is not the same as for closed strings, but they yield the same result of \({\cal A} \sim s^{2-\Delta/2}\).

One can make here the following observation. The closed string amplitude behaves as \(s e^{-\beta}\) at the fixed angle, high energy limit, and the open string one as \(e^{-\beta}\) without the leading factor of \(s\), but the process of scaling \(s \to s/r^2\) and performing the integral leads to the same result.  In fact, one can start with \({\cal A}_{10} \sim s^n e^{-\beta}\) for any power \(n\), and after integration get that the behavior is \(s^n \beta^{2-n+\Delta/2} \sim s^{2-\Delta/2}\).

\subsubsection{In the soft wall model}
\begin{figure}[h!]
\centering
\includegraphics[scale=0.5]{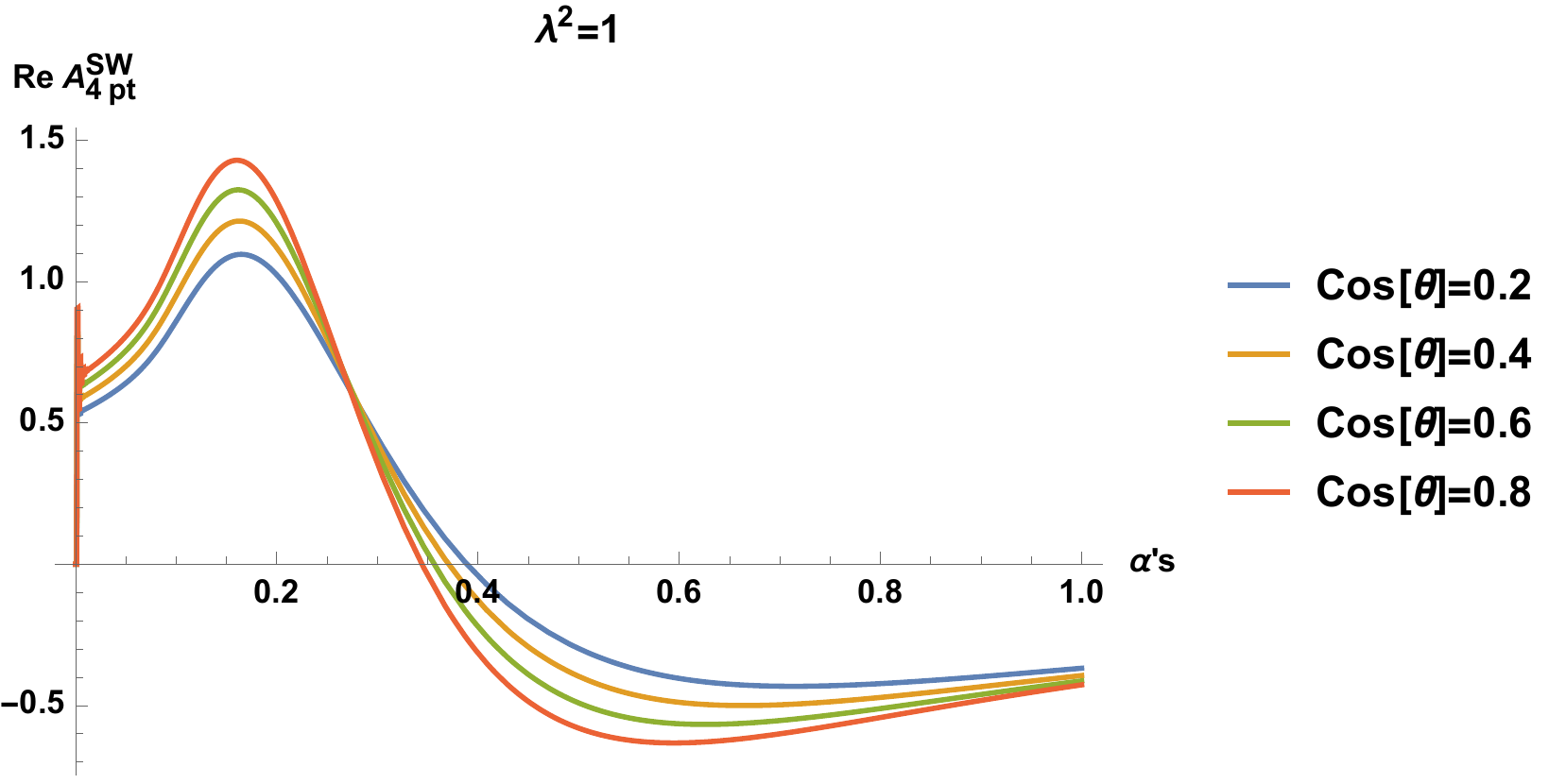} \includegraphics[scale=0.5]{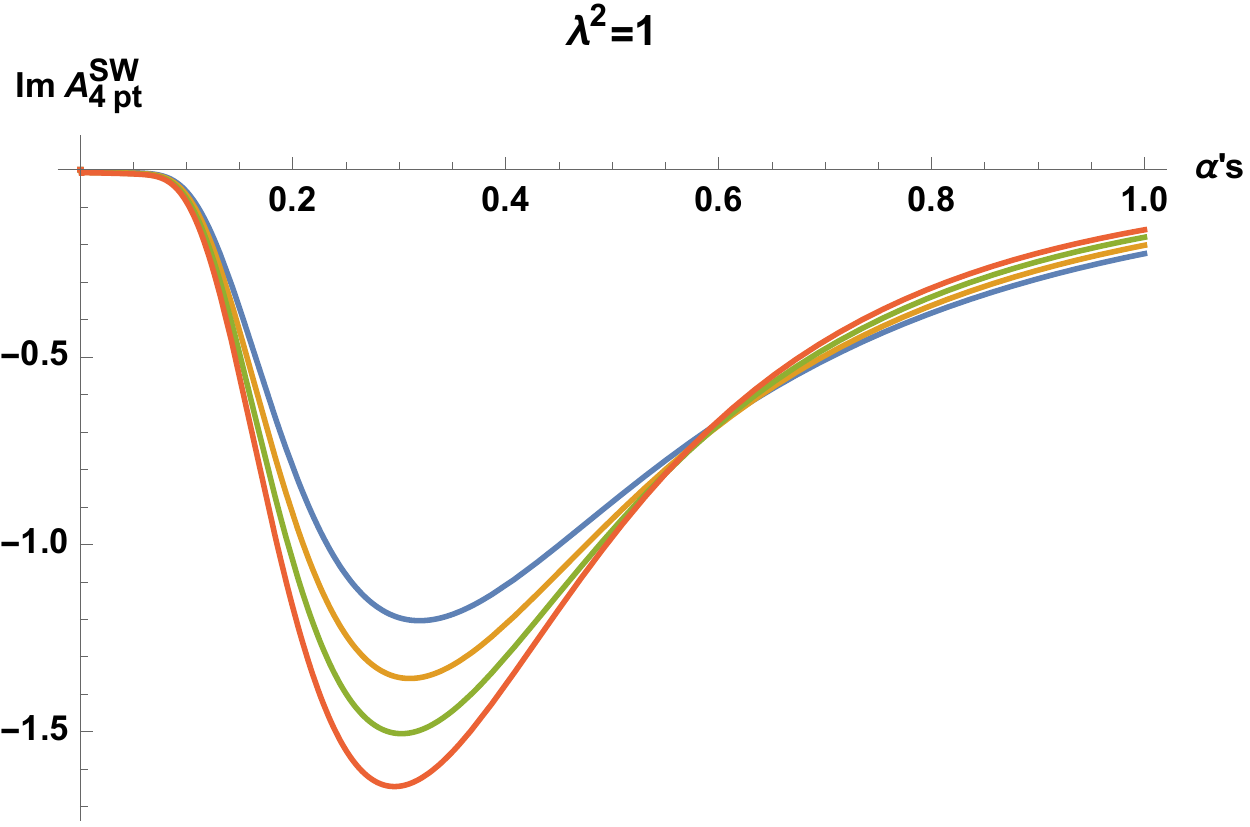}
\caption{\label{fig:open_super_sw} Real and imaginary parts of the open string amplitude in the soft wall at different values of $cos\theta$.}
\end{figure}

In the soft wall background the amplitude, written in the pole expansion, after integration over the radial coordinate, turns out to be 
\be
\begin{split}
{\cal A}_{4}^{SW}=\sum_{n=0}^{\infty} {(-1)^{n{+}1}\over (n+1)!} {\Gamma\big(c_{+}(n{+}1)+1\big)\over \Gamma\big(c_{+}(n{+}1)-n\big)}&\exp\left(-{\lambda^{2}(n{+}1)\over \alp s}\right) {\alp s\over 2 (n{+}1)^{2}}  \left(-{\alp s \over n{+}1}\right)^{1{-}{\Delta\over 2}} \\
 &\Gamma\left({\Delta\over 2}{-}2\right)\widetilde\Gamma\left(3{-}{\Delta\over 2}\,;-{\lambda^{2}(n{+}1)\over \alp s}\right)
 \end{split}
\ee
 The behavior of the amplitude for $\Delta=8$ is shown in figure \ref{fig:open_super_sw}.

The fixed angle limit in the soft wall yields
\be
{\cal A}^{\text{FA}}_{4}=-{s\over t} \int_{0}^{\infty} dz\, z^{\Delta{-}5} e^{-{(\beta+\lambda^{2}) z^{2}}}={1\over 2 c_{+}} (\beta_{st}+\lambda^{2})^{2-{\Delta\over 2}} \Gamma\left({\Delta\over 2}-2\right) \sim s^{2-\frac\Delta2}
\ee

\subsection{Veneziano amplitude}
Barring the prefactor $\epsilon^{ijk} \varepsilon^{\mu\nu\rho\sigma} \omega_\mu k^1_\nu k^2_\rho k^3_\sigma$, accounting for flavor symmetry and kinematics, the Veneziano amplitude \cite{Veneziano:1968yb} is Euler's Beta function
\be {\cal A}_V(s,t) = \frac{\Gamma(-\alpha(s))\Gamma(-\alpha(t))}{\Gamma(-\alpha(s)-\alpha(t))} \ee
where we take linear trajectories
\be \alpha(s) = \alp s + \alpha_0 \ee
The intercept is \(\alpha_0 = 1\) for the bosonic string tachyon amplitude, but one can take some generic value in the following. For unitarity of the original (non-holographic) amplitude we should require \(\alpha_0 \leq 1\) (see for instance \cite{Cappelli:2012cto}).

The expansion of the Veneziano amplitude in the \(s\)-channel has a simple form,
\be {\cal A}_V(s,t) = \sum_{n=0}^\infty \frac{(\alpha(t)+1)(\alpha(t)+2)\ldots(\alpha(t)+n)}{n!} \frac{1}{\alpha(s)-n} \ee
while the fixed angle, high energy limit can be written as
\be {\cal A}_V^{\text{FA}}(s,t) \simeq e^{-f(\theta)\alpha(s)} \label{eq:A_Ven_FA}\ee
with
\be f(\theta) = -c_- \log c_- - c_+\log c_+ \ee

\subsubsection{In the hard wall model}
After integration over the radial coordinate in the hard wall model the amplitude becomes
\be {\cal A}^{\text{HW}}_4(s,t) = \sum_{n=0}^\infty \frac{{\cal R}_n(\theta)}{(\Delta-4)(n-\alpha_0)} {}_2 F_1(1,\frac{\Delta}{2}-2,\frac\Delta2-1,\frac{\alp  s}{n-\alpha_0}) \ee
where
\be {\cal R}_n(\theta) = \frac{(-1)^n}{n!}\frac{\Gamma[(n-\alpha_0)c_- - \alpha_0]}{\Gamma[(n-\alpha_0)c_- - n-\alpha_0]} \ee
is a polynomial of degree \(n\) in \(\cos(\theta)\).

Starting from the high energy, fixed angle limit of the Veneziano amplitude \eqref{eq:A_Ven_FA}, we find the expression
\begin{align} A(s,t) &= \int_{r_0}^\infty  dr r^{3-\Delta} \exp\left(- f(\theta) \alp s/r^2\right) \\
&= \frac12 e^{-\alpha_0 f(\theta)} \left[\Gamma\left(\frac{\Delta-4}{2}\right)-\Gamma\left(\frac{\Delta-4}{2},f(\theta) \alp  s\right)\right] \left( \alp  s f(\theta)\right)^{2-\frac\Delta2} \sim s^{2-\frac\Delta2}\end{align}

\subsubsection{In the soft wall model}
In the soft wall background the Veneziano amplitude becomes
\be
\begin{split}
{\cal A}_{4}^{SW}(s,t)=\sum_{n=0}^{\infty} \frac{(-1)^n}{n!}\frac{\Gamma[(n+\alpha_0)c_- - \alpha_0]}{\Gamma[-(n+\alpha_0)(1-c_-)]} &\exp\left(-{\lambda^{2}(n{+}1/2)\over \alp s}\right) {\alp s\over 2 (n{+}1/2)^{2}}  \left(-{\alp s \over n{+}1/2}\right)^{1{-}{\Delta\over 2}} \\
 &\Gamma\left({\Delta\over 2}{-}2\right)\widetilde\Gamma\left(3{-}{\Delta\over 2}\,;-{\lambda^{2}(n{+}1/2)\over \alp s}\right)
 \end{split}
\ee

\subsection{Lovelace-Shapiro amplitude}
We now pass to consider the Lovelace-Shapiro (LS) amplitude \cite{Lovelace:1968kjy, Shapiro:1969km} that was proposed soon after Veneziano's breakthrough paper \cite{Veneziano:1968yb}. LS has the virtue of  exposing the Adler zeroes (vanishing for zero momenta of the external `pions'), as expected for (pseudo)Goldstone bosons like pions. Another nice property of LS amplitude is the presence of two Regge trajectories with the same slope and different intercept. The pion trajectory $\alpha' s = n$ and the $\rho$-meson trajectory\footnote{The LS model has no massless vector boson in the spectrum.} 
$\alpha' s = n+ {1\over 2}$. The former yields `massless' pions at $n=0$, the latter a massive $\rho$ meson with a mass $m^2_\rho = 1/2\alpha'$ that fits the experimental value $m_\rho\approx 770 MeV$ for $\alpha' \approx 1 GeV^{-2} \approx 1/m_{p}^2$ (inverse square proton mass).

In hindsight the LS amplitude may be viewed as the 4-point amplitude of Neveu-Schwarz tachyons with a special choice of their `internal' momenta ${Q}_i$ that makes them massless and shifts the ground state of the intermediate $s, t$ channels \cite{Bianchi:2020cfc}. For this reason the flavor-ordered LS amplitude is `tree level' unitary, in the sense of positive residues at the poles.

The flavor-ordered Lovelace-Shapiro amplitude proposed for pion-pion scattering is
\be
{\cal A}_{LS}(s,t)=C_{4pt}\,{\Gamma\left({1\over 2}-\alp s \right)\Gamma\left({1\over 2}-\alp t \right) \over \Gamma\left(-\alp s -\alp t \right)}
\ee
where $C_{4pt}$ incorporates the flavor symmetry \`a la Chan-Paton.

The high energy behavior is the same as for the amplitudes considered previously. In the following we write the expressions in the pole expansion only.

\subsubsection{In the hard wall model}
Taking steps similar to the previous sections one finds
 \be
 {\cal A}^{HW}_{LS}={1\over (4-\Delta)}\sum_{n=0}^{\infty} {(-1)^{n{+}1}\over (n+1)!} {\Gamma\left({1\over 2} + c_{+}(n{+}1/2)\right)\over \Gamma\big(-c_{-}(n{+}1/2)\big)}{\,_{2}F_{1}\Big(1\,; {\Delta\over 2}-2\,;{\Delta\over 2}-1\,; {\alp  s\over n{+}1/2}  \Big)}
 \ee
 
\begin{figure}[h!]
 \centering
 \includegraphics[scale=0.5]{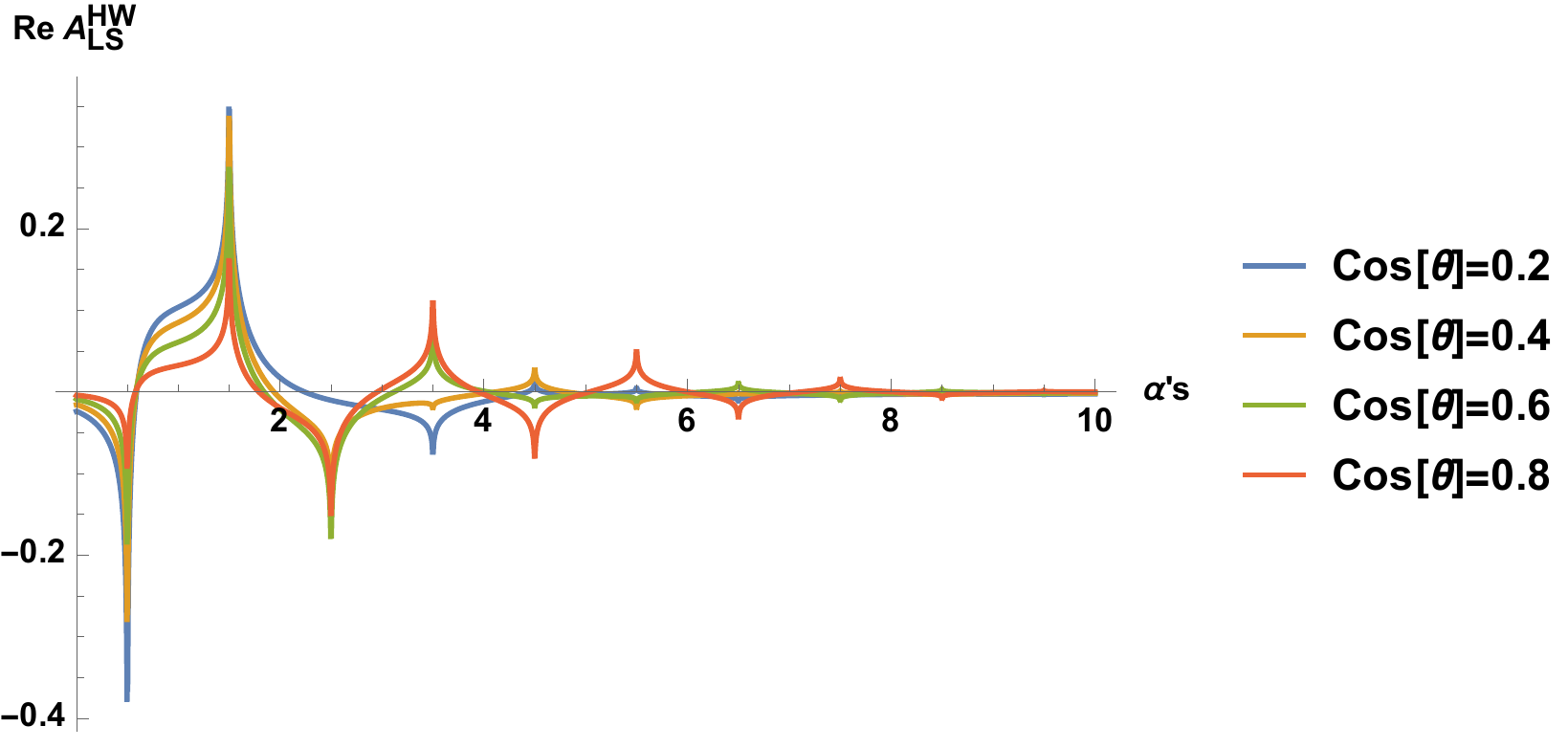}  \includegraphics[scale=0.5]{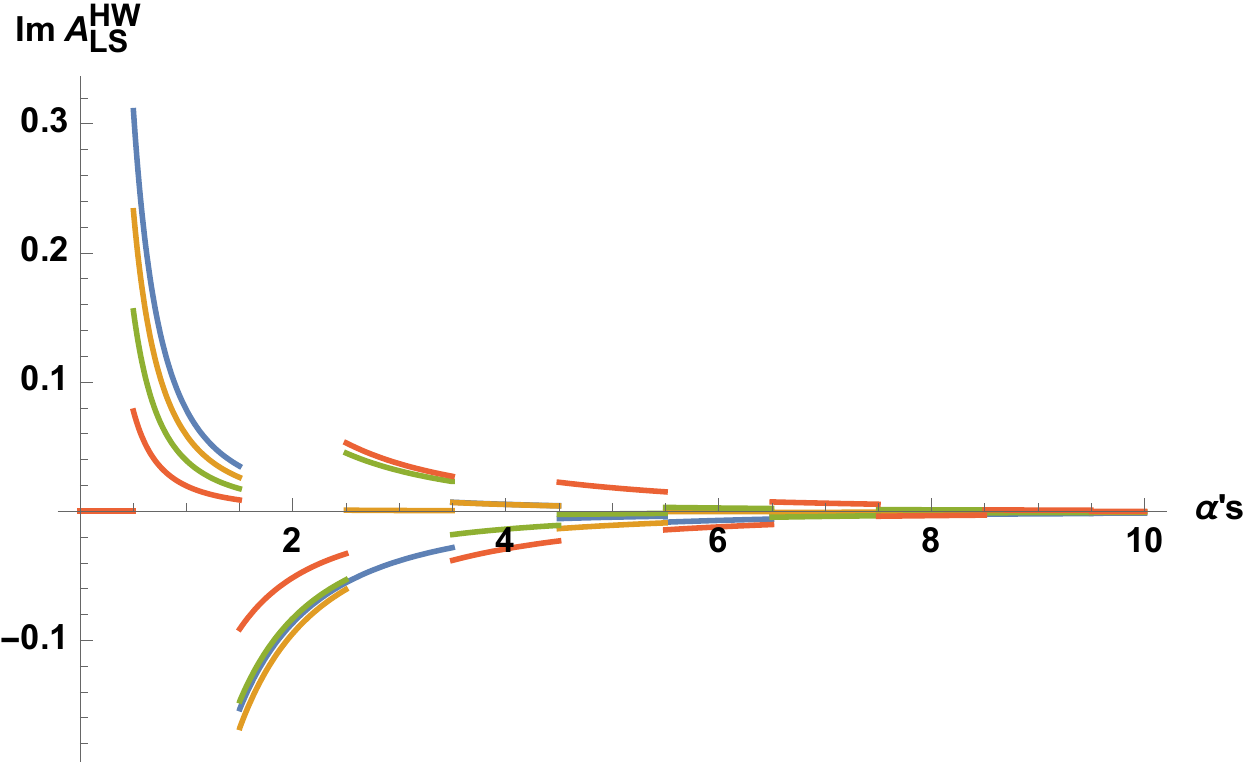}
 \caption{\label{fig:open_LS_hw} Real and imaginary part of the LS amplitude in the hard wall background at different values of $c_{\theta_{\pm}}$ for $\Delta=8$}
 \end{figure}

 The amplitude with \(\Delta=8\) is plotted in figure \ref{fig:open_LS_hw}.

\subsubsection{In the soft wall model}
\begin{figure}[h!]
\centering
\includegraphics[scale=0.5]{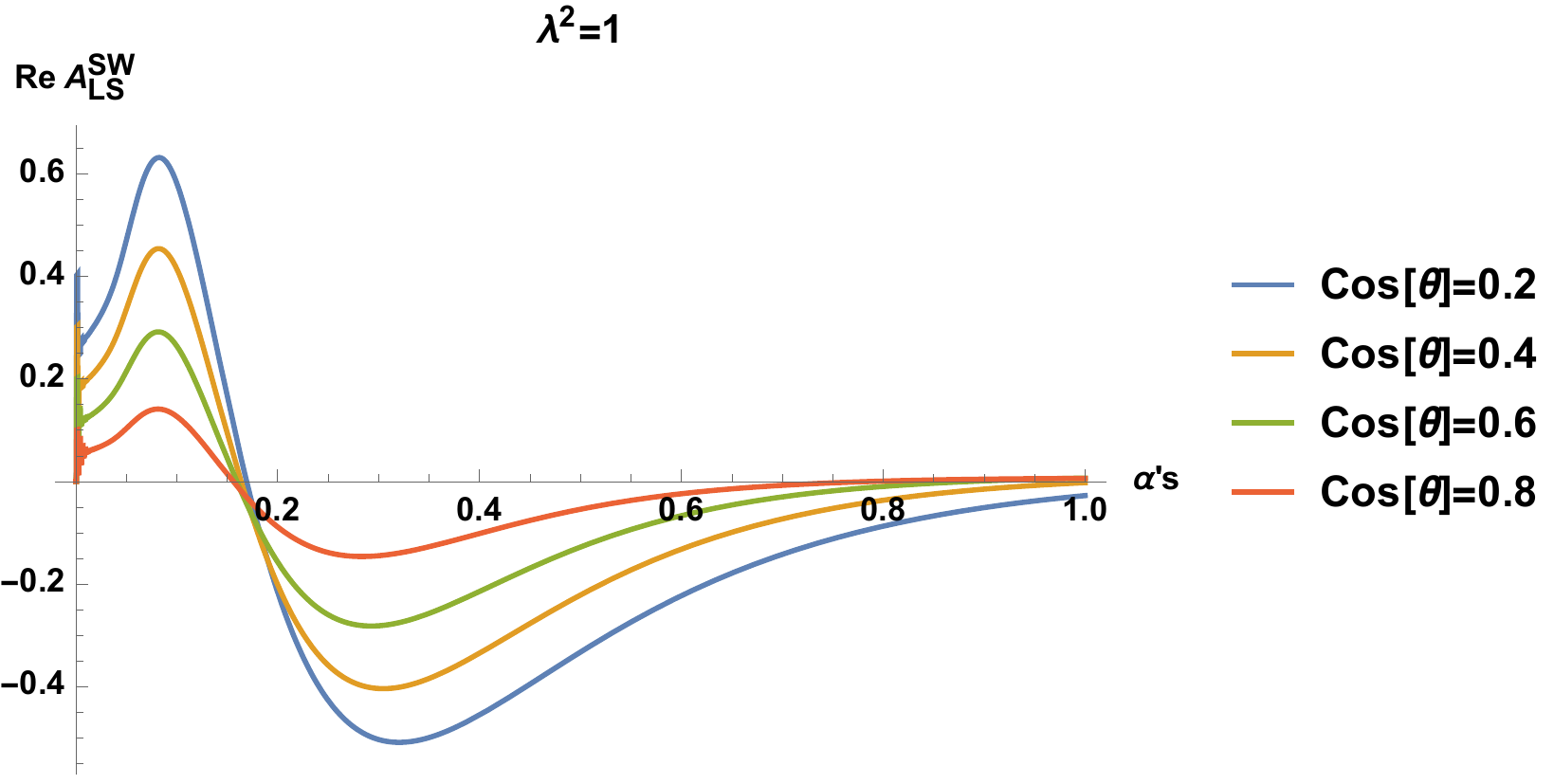} \includegraphics[scale=0.5]{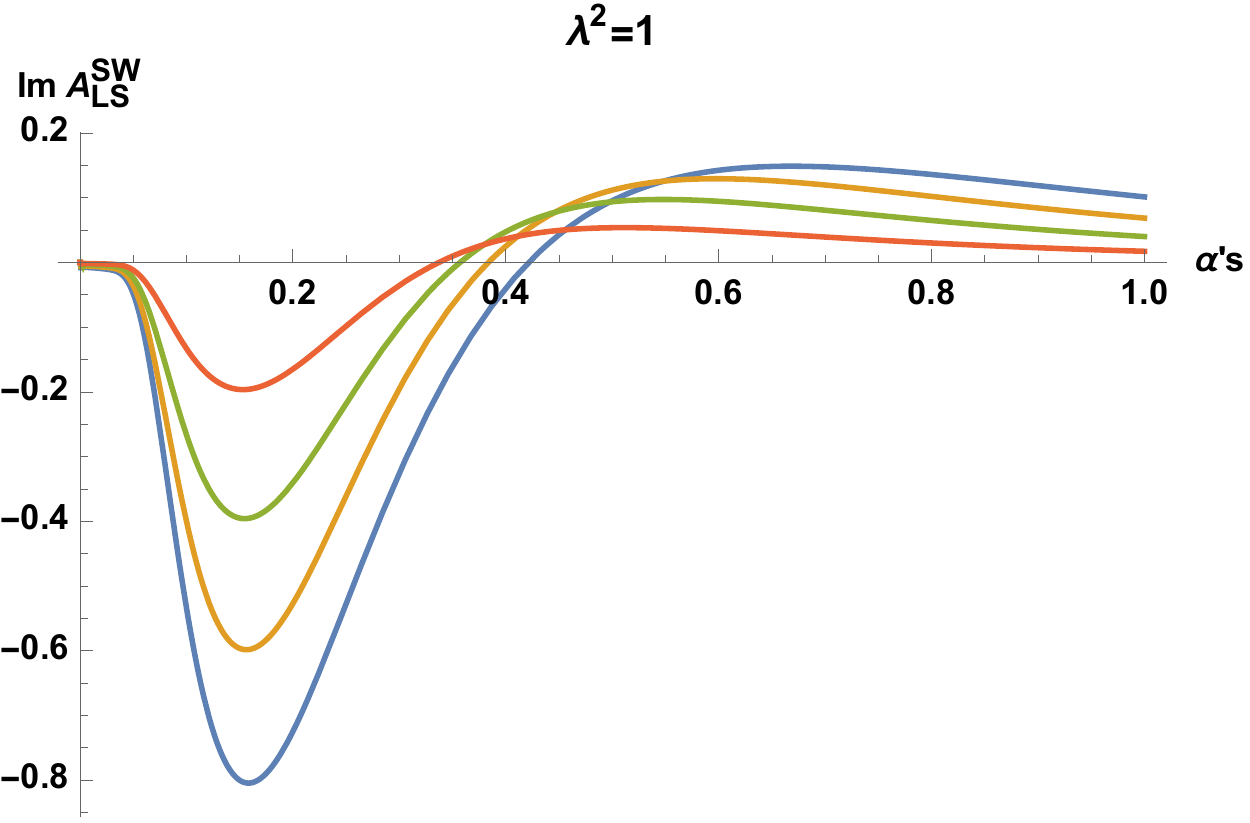}
\caption{Real and imaginary parts of the LS amplitude at different values of $cos\theta$ in the soft wall model.}
\label{SW_LS}
\end{figure}

In the SW background the LS amplitude is given by
\be
\begin{split}
{\cal A}_{4p}^{SW}=\sum_{n=0}^{\infty} {(-1)^{n{+}1}\over (n+1)!} {\Gamma\left({1\over 2} + c_{+}(n{+}1/2)\right)\over \Gamma\big(-c_{-}(n{+}1/2)\big)}&\exp\left(-{\lambda^{2}(n{+}1/2)\over \alp s}\right) {\alp s\over 2 (n{+}1/2)^{2}}  \left(-{\alp s \over n{+}1/2}\right)^{1{-}{\Delta\over 2}} \\
 &\Gamma\left({\Delta\over 2}{-}2\right)\widetilde\Gamma\left(3{-}{\Delta\over 2}\,;-{\lambda^{2}(n{+}1/2)\over \alp s}\right)
 \end{split}
\ee
The amplitude behavior for the case $\Delta=8$ is shown in figure \ref{SW_LS}.

%%%%%%%%%%%%%%%%%%%%%%%%%%%%%%%%%%

\subsection{Open string scattering in the Witten-Sakai-Sugimoto model}
To include open strings in the story one has to embed flavor branes in the ten dimensional background. This is true both in Witten's model and in bottom-up models like the hard and soft wall models we used above. In the background of Witten's model eq. \eqref{eq:backgroundWitten}, this embedding results in the Sakai-Sugimoto model. As was briefly described in section \ref{sec:quarks}, one embeds a stack of $N_f$ $D8$ branes and as many $\overline{D8}$-branes. These two stacks of brane merge together in the holographic region that corresponds to the IR in the dual field theory, thus providing a geometrical mechanism of chiral symmetry breaking.

From the point of view of computing the scattering amplitudes there is a major difference between the closed and open string cases. For the former the weight function was the wave function of the corresponding bulk modes. In the WSS model the usual wave functions are those associated with the fluctuations of the flavor branes for open strings that correspond to scalar/pseudoscalar mesons, or the fluctuations of the bulk gauge fields corresponding to vector mesons in the adjoint of the flavor group.\footnote{For reviews see \cite{Erdmenger:2007cm,Rebhan:2014rxa} or the book \cite{Ammon:2015wua}.}

The flavor branes in the Sakai-Sugimoto model are \(D8\) branes, and therefore their embedding is described by a function of \(U(x_4)\). This means that we generally need to consider wave functions which are functions of both coordinates \(\psi(U,x_4)\), that are centered around the classical configuration of the branes. In the integration of the amplitude in the Polchinski-Strassler approach we will need to integrate explicitly over \(x_4\) as well.

As a simplifying assumption we can consider only a simple power dependence on the radial coordinate, such that all effects considered, we can perform the integral with the same measure as before,
\be {\cal A}_4(s,t,u) = \int_{U_\Lambda}^\infty du\sqrt{-g}e^{-2\phi} U^{-\Delta} {\cal A}_{10}(\tilde s,\tilde t,\tilde u) \ee
for some value \(\Delta\). The scaling of the Mandelstam variables is the same as before, {\it i.e.} \(\tilde s = s U^{-3/2}\).

We now place the amplitude \eqref{eq:A_open_vector} in the 10D background of eq. \eqref{eq:backgroundWitten}. If we pursue the same approach, as a function of \(\Delta\) the results are as for the closed string in the 10D case (section \ref{sec:Witten10}). The fixed angle amplitude behaves as
\be {\cal A}_4^{\text{FA}} \sim s^{\frac73-\frac23\Delta} \label{eq:A_FA_WSS}\ee

Next one can try to evaluate \(\Delta\) based on what we know of the fluctuation modes corresponding to open strings.

According to the analysis of \cite{Sakai:2004cn} the pseudoscalar modes have wave functions that are asymptotically \(U^{-3}\) (the same as the scalar bulk modes), while the vector fluctuation modes have \(U^{-3/2}\). Substituting these values (times four) into the last equation,
\be {\cal A}_{\text{pseudo}}^{\text{FA}} \sim s^{-\frac{17}{3}} \ee
\be {\cal A}_{\text{vector}}^{\text{FA}} \sim s^{-\frac{11}{3}} \ee

The spectrum of scalar mesons in the Sakai-Sugimoto model was computed in \cite{Mintakevich:2008mm} by introducing fluctuations of the embedding of the flavor branes. The classical configuration of the \(D8\) branes is a function \(x_4(U)_{\text{cl}}\), to which one adds fluctuations,
\be x_4(U,x^\mu) = x_4(U)_{\text{cl}} + \sum_{n} e^{-i k_n \cdot x} \xi_n(U) \ee
such that \(k_n^2 = -m_n^2\). The eigenmodes obey the equation
\be \partial_U[U^{11/2}\gamma^{-3/2}\partial_U \xi_n] + m_n^2 R^3 U^{5/2}\xi_n = 0 \ee
where
\be \gamma(U) = \frac{U^8}{f(U)U^8-f(U_0)U_0^8} \ee
If we can identify the radial dependence of \(\xi_n(U)\) with \(\psi(U)\), we can read the value of \(\Delta\) from the large \(U\) regime. In that region, \(\gamma \approx 1\) and the equations simplifies. Then, one can show that at large \(U\),
\be \xi_n(U) \sim U^{-9/2} \ee
If we take \(\Delta = 4\times\frac92 = 18\), the prediction of \eqref{eq:A_FA_WSS} is that
\be {\cal A}_{\text{scalar}}^{\text{FA}} \sim s^{-\frac{29}3} \ee
It is not unlikely that in a more precise formulation where the dependence on \(x_4\) is considered the values of \(\Delta\) for the different modes, and therefore the power of \(s\) in the high energy behavior of the amplitudes will change.

Clearly the various results are unsatisfactory, as all are meant to describe processes with a total of eight partons, and therefore all should, according to the QCD expectation behave the same at high energy, specifically as \(s^{-2}\). Finding the correct prescription for open string scattering in the WSS model is an open problem.

%%%%%%%%%%%%%%%%%%%%%%%%%%%%%%%%%%%%%%%%%%%%%%%%%

\section{Conclusions and summary} \label{sec:summary}
The aim of the present investigation was extending and generalizing the approach to hadron scattering proposed by Polchinski and Strassler (PS) \cite{Polchinski:2001tt} to various different holographic models. Namely we used the hard wall (HW), soft wall (SW) and Witten-Sakai-Sugimoto (WSS) models. In addition to glueballs, dual to closed string states, we have also considered mesons, dual to open string states. More precisely we focused on light-light (pseudo-)scalar mesons made of light (massless) quarks. Further generalization to vector or higher-spin mesons or to light-heavy and heavy-heavy (pseudo-)scalar mesons is possible but beyond the scope of the present investigation. 

In all cases we have shown the hard partonic behavior at large energies and checked the analytic structure at low (or intermediate) energies. Notwithstanding the limitations of the approximation, that amounts to integrating a flat space string amplitude with rescaled momenta over the radial/holographic direction with `supergravity' wave functions, the results are encouraging.

Before drawing our conclusions and delineating directions for future investigation, let us schematically summarize and comment on the main outcomes of our present investigation, restating the results from section \ref{sec:closed}.

We calculated the amplitude in three methods/regimes, and several models:

\noindent 1) Fixed angle approximation ($|t|\sim s$) before radial/holographic integration:
\begin{itemize}
\item Hard wall (polynomial $\psi$): \({\cal A}\sim s^{2-\Delta/2}\). We can take \(\Delta = 8\), (replacing dimension with twist or number of constituents $\Delta_i\rightarrow 2$) so that \({\cal A} \sim s^{-2}\).
\item Hard wall (Bessel $\psi$): \({\cal A}\sim s^{2-\Delta/2}\). {\it Caveat}: for a massless scalar in \(AdS_5\) \(\Delta_i = 4\), so \(\Delta = 16\). Using `exact' $\psi$ corresponds to taking \(\Delta_i\), not the twist or the number of constituents (gluons) as desired, and therefore \({\cal A}\sim s^{-6}\).
\item Soft wall: \({\cal A}\sim s^{2-\Delta/2}\). We can take \(\Delta = 8\) (replacing dimension with twist or number of constituents) so that \({\cal A} \sim s^{-2}\).
\item Witten's model (11D): \({\cal A} \sim s^{3-\Delta/2}\). Here the dependence on \(\Delta\) is altered because asymptotically the background is a higher dimensional \(AdS_7\). Then the results do not match with the expectation in four dimensional QCD. In addition, from the scalar wave function we read \(\Delta = 24\) for a final result of \({\cal A} \sim s^{-9}\).
\item Witten's model (10D): \({\cal A} \sim s^{7/3-2\Delta/3}\). The warp factor of \(U^{3/2}\) in this background results in fractional powers rather than integer ones, at variant with the 11D formulation. With \(\Delta=12\) we find \({\cal A} \sim s^{-17/3}\).
\end{itemize}

\noindent 2) Pole expansion, integration, summation:
\begin{itemize}
\item All models: \(\Rea[{\cal A}] \sim s^{-1}\), \(\Imm[{\cal A}] \sim s^{\alpha}\). Where \(\alpha\) for each model is the same power that one finds in the fixed angle calculation detailed above (\(\alpha = 2-\Delta/2\) for hard wall, etc.). With the exception of:
\item Soft wall model: \(\Rea[{\cal A}] \sim s^{3-\Delta/2}\), \(\Imm[{\cal A}] \sim s^{2-\Delta/2}\).
\end{itemize}

\noindent  3) Regge regime, large $s$, fixed $t$ (small angle):
\begin{itemize}
\item All models:  $ \Rea{\cal A}\sim {s^{2}}\,t^{-2+\alpha}\, (\log{s\over t})^{-1+\alpha}$; $ \Imm{\cal A}\sim {s^{2}}\,t^{-2+\alpha} \,(\log{s\over t})^{-2+\alpha}$
with the same \(\alpha\) as above.
\end{itemize}

In sections \ref{sec:general_background} and \ref{sec:general_background_closed} we wrote general expressions that depend on the parameters of the chosen background.

The results are somewhat model dependent and suggest the following interpretation. First of all, we find substantial agreement between the two bottom-up models, namely HW and SW, in the different (high-energy) regimes and in the various approaches we pursued to arrive at the final results. Some discrepancies that we find among these and the results for the top-down WSS model can be imputed to the higher-dimensional origin of the latter that, despite being a bona fide solution of the low-energy supergravity emerging from string theory or M-theory, generates unavoidable contamination with the KK modes, whose decoupling is one of the major challenges towards a reliable microscopic description of the QCD string.  Backgrounds that are asymptotically $AdS_5$ generate the correct power dependence on the energy, $s^{2-\Delta/2}$, if one can associate $\Delta$ with the number of partons.

In section \ref{sec:open} we checked the same models starting from open string amplitudes and confirmed that they have the same behavior. A more thorough and precise analysis taking into account the presence of flavor branes is necessary there, but is left as a subject for future work.

In addition to the high energy behavior, we have found a somewhat puzzling structure of logarithmic singularities at low energies that replace the poles in the original flat space amplitude.  This in an immediate consequence of the integration over the radial coordinate. At fixed angle, the (logarithmic) singularities become less pronounced as we increase the energy. This can be interpreted as an effect of asymptotic freedom. One could argue that the presence of logarithmic singularities rather than poles is a problem of the present analysis. It is an open question whether in holographic models not considered here the simple pole structure can be preserved. In order to have only poles the holographic direction should be effectively discretized as in \cite{Andreev:2004sy, Veneziano:2017cks}  but this would raise subtle issues with factorization \cite{Bianchi:2020cfc}. 

In this paper we took one further step in the path paved in \cite{Polchinski:2001tt,Polchinski:2002jw}. There are obviously many open questions that deserve further study. Let us mention some of them:
\begin{itemize}
\item
In a similar way to identifying a confining background using the stringy Wilson line \cite{Kinar:1998vq} one may wonder whether one can characterize backgrounds that ensure the transition from soft stringy to hard partonic scattering amplitudes. 
\item 
In the present work we looked at backgrounds where the scaling of the Mandelstam variables is with a simple power in the radial variable, e.g. \(\tilde s = s/r^2\) as in $AdS$. In some holographic models, e.g. Klebanov-Strassler \cite{Klebanov:2000hb}, this is not the case, so one can extend our analysis to a more general case. In particular one could check if in that model, there are also logarithmic singularities after integration.
\item
The various different models studied in this work yielded different results for the amplitudes in particular at fixed angle. Understanding these differences and evaluating the quality of the various results is still an open question.
\item
The prescription used in this paper, following the seminal paper of \cite{Polchinski:2001tt} is a hybrid formulation. Devising a procedure fully based on string theory and not a hybrid one would be an important challenge.
\item
The HISH program maps the holographic stringy setup into strings in flat four dimensions. In the same spirit one would like to find a prescription in four dimension that could reproduce the transition between soft to hard scattering. The integration over the holographic coordinate with a weight function may be associated with averaging of momentum dependent string tension.
\item 
Perhaps the most interesting process that one can imagine getting from the stringy holographic picture is proton-proton scattering. For that purpose  one has to use in the hybrid procedure the wave functions associated with the  baryons which takes the form of flavor instantons in the model of \cite{Kaplunovsky:2015zsa}. The instanton is a soliton of the five dimensional theory on the flavor brane after integrating over the four sphere. It will be interesting to analyze this system and to check to what extent it reproduces the known proton-proton scattering amplitude.
\end{itemize}

Although a full-fledged string description of QCD might still be hard to achieve we would like to comment on other less phenomenological and more `stringy' approaches and how they can be related to the one introduced by PS and adopted here.

\begin{itemize}
\item
Following Armoni and Ireson \cite{Armoni:2016nzm}, one can start from WSS model, expand the background around flat space in order to account for the world-sheet dynamics and compute loop corrections in $\alp /R^2$ to Veneziano amplitude. Quite remarkably one finds some promising hints to the bending of Regge trajectories in the physical domain  ($t<0$), as expected from asymptotic freedom and the ensuing hard/partonic behavior of QCD scattering amplitude.

\item
Following e.g. \cite{Armoni:2009vv}, one can rely on the equivalence between non-critical superstrings and supestrings on ${\bf R}^4 \times SL(2)/U(1)$ in order to derive the closed string spectrum, vertex operators for low-lying states, $D$-branes and boundary states for a stringy version of ${\cal N}=1$ supersymmetric Yang-Mills theory (SYM). The cigar geometry nicely accounts for the breaking of the $U(1)$ R-symmetry in the IR, and D-branes with various embeddings provide a compelling description of flavor. Yet computing one- and two-point correlators and decay rates (3-point amplitudes) looks extremely laborious, let alone binary scattering amplitudes, such as the ones considered in the present investigation. 

\item 
Following Andreev and Siegel \cite{Andreev:2004sy} and Onofri, Veneziano and Yankielowicz \cite{Veneziano:2017cks}, one can generalize binary scattering for mesons/pions \`a la Lovelace-Shapiro or \`a la Veneziano by summing (rather than integrating) over replicas of the Regge trajectories with different slopes.
Unitarity requires the poles to be a subset of those of the original amplitude. In particular ${\alp \over k} s = n$ for integer $k$ seems to work.  This means ${\alp_k} s = n$ or ${\alp_k} = {\alp \over k}$. 

In \cite{Andreev:2004sy, Veneziano:2017cks} it is shown that weighting the contributions with $w_k^{(N)}\sim g_k/k^N$ gives `reasonable' 4-point amplitudes with Regge behavior in the unphysical positive $t$ regime and hard/partonic behavior in the physical negative $t$ regime.  The problem is to go beyond 4-points and check factorization. This looks problematic, as already observed in \cite{Bianchi:2020cfc}, since one has a single sum over $k$, while one would naively expect one for each sub-amplitude. Only in the large $k$ continuum limit holographic formulae \`a la Polchinski-Strassler seem to produce the correct factorization as we found above. Note that \(\alp_k\) is the expected effective slope of \(k\)-folded strings, though the explicit relation is not made in the papers cited. An exact analysis of the contribution of folded string configurations to scattering amplitudes could be beneficial.

\item  Let us comment on the inclusion of higher-loop contributions to the string amplitude. In a series of papers, using saddle-point method to integrate over the moduli space of punctured Riemann surfaces, Gross, Mende and Ooguri \cite{Gross:1987ar,Gross:1987kza,Mende:1989wt} estimated a correction to the high-energy behavior of 4-point amplitudes to be of the form
\[
\alp s f(\theta) \rightarrow \sqrt{\alp s} f(\theta) + ...
\]
It is worth noticing that, although integration over the radial variable and integration over moduli (especially in a saddle-point approximation) may not commute with one another, a drastic change as the above would affect most of the conclusions reached in the PS approach. Clearly it would be crucial to check whether loop-corrections to {\it full-fledged} holographic amplitudes would confirm or disprove GMO saddle-point estimates but this is well beyond our present abilities.  

\end{itemize}

\section*{Acknowledgments}

We would like to thank A.~Aldi, S. Caron-Huot,  D.~Consoli, A.~Guerrieri, P.~Di~Vecchia, F.~Fucito, Z.~ Komargodski, J.~ Leutgeb, A.~ Rebhan, A.~ Sever, M.~ Strassler and S. Yankielowicz for useful discussions.
The work of M.~B and M.~F was partially supported by the excellence grant 'Beyond borders' of the University of Roma Tor Vergata and by the MIUR PRIN Grant 2020KR4KN2 ‘String Theory as a bridge between Gauge Theories and Quantum Gravity’.
J.~S. was supported by a center of excellence of the Israel Science Foundation (grant number 2289/18).  D.~W. is supported by the Quantum Gravity Unit at the Okinawa Institute of Science and Technology Graduate University.
J.~ S. would like to thank NYU for their hospitality during a period when part of the research work has been conducted.

\clearpage
\bibliographystyle{JHEP}
\bibliography{SACS}

\providecommand{\href}[2]{#2}\begingroup\raggedright\begin{thebibliography}{10}

\bibitem{Polchinski:2001tt}
J.~Polchinski and M.J.~Strassler, \emph{{Hard scattering and gauge / string
  duality}}, \href{https://doi.org/10.1103/PhysRevLett.88.031601}{\emph{Phys.
  Rev. Lett.} {\bfseries 88} (2002) 031601}
  [\href{https://arxiv.org/abs/hep-th/0109174}{{\ttfamily hep-th/0109174}}].

\bibitem{Veneziano:1968yb}
G.~Veneziano, \emph{{Construction of a crossing - symmetric, Regge behaved
  amplitude for linearly rising trajectories}},
  \href{https://doi.org/10.1007/BF02824451}{\emph{Nuovo Cim.} {\bfseries A57}
  (1968) 190}.

\bibitem{Virasoro:1969me}
M.A.~Virasoro, \emph{{Alternative constructions of crossing-symmetric
  amplitudes with regge behavior}},
  \href{https://doi.org/10.1103/PhysRev.177.2309}{\emph{Phys. Rev.} {\bfseries
  177} (1969) 2309}.

\bibitem{Green:1982sw}
M.B.~Green, J.H.~Schwarz and L.~Brink, \emph{{N=4 Yang-Mills and N=8
  Supergravity as Limits of String Theories}},
  \href{https://doi.org/10.1016/0550-3213(82)90336-4}{\emph{Nucl. Phys. B}
  {\bfseries 198} (1982) 474}.

\bibitem{Sonnenschein:2014jwa}
J.~Sonnenschein and D.~Weissman, \emph{{Rotating strings confronting PDG
  mesons}}, \href{https://doi.org/10.1007/JHEP08(2014)013}{\emph{JHEP}
  {\bfseries 1408} (2014) 013}
  [\href{https://arxiv.org/abs/1402.5603}{{\ttfamily 1402.5603}}].

\bibitem{Polchinski:2002jw}
J.~Polchinski and M.J.~Strassler, \emph{{Deep inelastic scattering and gauge /
  string duality}},
  \href{https://doi.org/10.1088/1126-6708/2003/05/012}{\emph{JHEP} {\bfseries
  05} (2003) 012} [\href{https://arxiv.org/abs/hep-th/0209211}{{\ttfamily
  hep-th/0209211}}].

\bibitem{Brower:2006ea}
R.C.~Brower, J.~Polchinski, M.J.~Strassler and C.-I.~Tan, \emph{{The Pomeron
  and gauge/string duality}},
  \href{https://doi.org/10.1088/1126-6708/2007/12/005}{\emph{JHEP} {\bfseries
  12} (2007) 005} [\href{https://arxiv.org/abs/hep-th/0603115}{{\ttfamily
  hep-th/0603115}}].

\bibitem{Erdmenger:2007cm}
J.~Erdmenger, N.~Evans, I.~Kirsch and E.~Threlfall, \emph{{Mesons in
  Gauge/Gravity Duals - A Review}},
  \href{https://doi.org/10.1140/epja/i2007-10540-1}{\emph{Eur. Phys. J.}
  {\bfseries A35} (2008) 81} [\href{https://arxiv.org/abs/0711.4467}{{\ttfamily
  0711.4467}}].

\bibitem{Karch:2006pv}
A.~Karch, E.~Katz, D.T.~Son and M.A.~Stephanov, \emph{{Linear confinement and
  AdS/QCD}}, \href{https://doi.org/10.1103/PhysRevD.74.015005}{\emph{Phys. Rev.
  D} {\bfseries 74} (2006) 015005}
  [\href{https://arxiv.org/abs/hep-ph/0602229}{{\ttfamily hep-ph/0602229}}].

\bibitem{Frolov:2002av}
S.~Frolov and A.A.~Tseytlin, \emph{{Semiclassical quantization of rotating
  superstring in AdS(5) x S**5}},
  \href{https://doi.org/10.1088/1126-6708/2002/06/007}{\emph{JHEP} {\bfseries
  06} (2002) 007} [\href{https://arxiv.org/abs/hep-th/0204226}{{\ttfamily
  hep-th/0204226}}].

\bibitem{Sonnenschein:2016pim}
J.~Sonnenschein, \emph{{Holography Inspired Stringy Hadrons}},
  \href{https://doi.org/10.1016/j.ppnp.2016.06.005}{\emph{Prog. Part. Nucl.
  Phys.} {\bfseries 92} (2017) 1}
  [\href{https://arxiv.org/abs/1602.00704}{{\ttfamily 1602.00704}}].

\bibitem{PandoZayas:2003yb}
L.A.~Pando~Zayas, J.~Sonnenschein and D.~Vaman, \emph{{Regge trajectories
  revisited in the gauge / string correspondence}},
  \href{https://doi.org/10.1016/j.nuclphysb.2003.12.006}{\emph{Nucl.Phys.}
  {\bfseries B682} (2004) 3}
  [\href{https://arxiv.org/abs/hep-th/0311190}{{\ttfamily hep-th/0311190}}].

\bibitem{Sonnenschein:2014bia}
J.~Sonnenschein and D.~Weissman, \emph{{A rotating string model versus baryon
  spectra}}, \href{https://doi.org/10.1007/JHEP02(2015)147}{\emph{JHEP}
  {\bfseries 1502} (2015) 147}
  [\href{https://arxiv.org/abs/1408.0763}{{\ttfamily 1408.0763}}].

\bibitem{Sonnenschein:2018fph}
J.~Sonnenschein and D.~Weissman, \emph{{Excited mesons, baryons, glueballs and
  tetraquarks: Predictions of the Holography Inspired Stringy Hadron model}},
  \href{https://doi.org/10.1140/epjc/s10052-019-6828-y}{\emph{Eur. Phys. J. C}
  {\bfseries 79} (2019) 326}
  [\href{https://arxiv.org/abs/1812.01619}{{\ttfamily 1812.01619}}].

\bibitem{Witten:1998qj}
E.~Witten, \emph{{Anti-de Sitter space and holography}}, {\emph{Adv. Theor.
  Math. Phys.} {\bfseries 2} (1998) 253}
  [\href{https://arxiv.org/abs/hep-th/9802150}{{\ttfamily hep-th/9802150}}].

\bibitem{Sakai:2004cn}
T.~Sakai and S.~Sugimoto, \emph{{Low energy hadron physics in holographic
  QCD}}, \href{https://doi.org/10.1143/PTP.113.843}{\emph{Prog. Theor. Phys.}
  {\bfseries 113} (2005) 843}
  [\href{https://arxiv.org/abs/hep-th/0412141}{{\ttfamily hep-th/0412141}}].

\bibitem{Brodsky:1973kr}
S.J.~Brodsky and G.R.~Farrar, \emph{{Scaling Laws at Large Transverse
  Momentum}}, \href{https://doi.org/10.1103/PhysRevLett.31.1153}{\emph{Phys.
  Rev. Lett.} {\bfseries 31} (1973) 1153}.

\bibitem{Brodsky:1974vy}
S.J.~Brodsky and G.R.~Farrar, \emph{{Scaling Laws for Large Momentum Transfer
  Processes}}, \href{https://doi.org/10.1103/PhysRevD.11.1309}{\emph{Phys. Rev.
  D} {\bfseries 11} (1975) 1309}.

\bibitem{Veneziano:2017cks}
G.~Veneziano, S.~Yankielowicz and E.~Onofri, \emph{{A model for pion-pion
  scattering in large-N QCD}},
  \href{https://doi.org/10.1007/JHEP04(2017)151}{\emph{JHEP} {\bfseries 04}
  (2017) 151} [\href{https://arxiv.org/abs/1701.06315}{{\ttfamily
  1701.06315}}].

\bibitem{Lovelace:1968kjy}
C.~Lovelace, \emph{{A novel application of regge trajectories}},
  \href{https://doi.org/10.1016/0370-2693(68)90255-4}{\emph{Phys. Lett. B}
  {\bfseries 28} (1968) 264}.

\bibitem{Shapiro:1969km}
J.A.~Shapiro, \emph{{Narrow-resonance model with regge behavior for pi pi
  scattering}}, \href{https://doi.org/10.1103/PhysRev.179.1345}{\emph{Phys.
  Rev.} {\bfseries 179} (1969) 1345}.

\bibitem{Andreev:2004sy}
O.~Andreev and W.~Siegel, \emph{{Quantized tension: Stringy amplitudes with
  Regge poles and parton behavior}},
  \href{https://doi.org/10.1103/PhysRevD.71.086001}{\emph{Phys. Rev. D}
  {\bfseries 71} (2005) 086001}
  [\href{https://arxiv.org/abs/hep-th/0410131}{{\ttfamily hep-th/0410131}}].

\bibitem{Caron-Huot:2016icg}
S.~Caron-Huot, Z.~Komargodski, A.~Sever and A.~Zhiboedov, \emph{{Strings from
  Massive Higher Spins: The Asymptotic Uniqueness of the Veneziano Amplitude}},
   \href{https://arxiv.org/abs/1607.04253}{{\ttfamily 1607.04253}}.

\bibitem{Sever:2017ylk}
A.~Sever and A.~Zhiboedov, \emph{{On Fine Structure of Strings: The Universal
  Correction to the Veneziano Amplitude}},
  \href{https://doi.org/10.1007/JHEP06(2018)054}{\emph{JHEP} {\bfseries 06}
  (2018) 054} [\href{https://arxiv.org/abs/1707.05270}{{\ttfamily
  1707.05270}}].

\bibitem{Kang:2004jd}
K.~Kang and H.~Nastase, \emph{{High energy QCD from Planckian scattering in AdS
  and the Froissart bound}},
  \href{https://doi.org/10.1103/PhysRevD.72.106003}{\emph{Phys. Rev. D}
  {\bfseries 72} (2005) 106003}
  [\href{https://arxiv.org/abs/hep-th/0410173}{{\ttfamily hep-th/0410173}}].

\bibitem{Hatta:2007he}
Y.~Hatta, E.~Iancu and A.H.~Mueller, \emph{{Deep inelastic scattering at strong
  coupling from gauge/string duality: The Saturation line}},
  \href{https://doi.org/10.1088/1126-6708/2008/01/026}{\emph{JHEP} {\bfseries
  01} (2008) 026} [\href{https://arxiv.org/abs/0710.2148}{{\ttfamily
  0710.2148}}].

\bibitem{Pire:2008zf}
B.~Pire, C.~Roiesnel, L.~Szymanowski and S.~Wallon, \emph{{On AdS/QCD
  correspondence and the partonic picture of deep inelastic scattering}},
  \href{https://doi.org/10.1016/j.physletb.2008.10.026}{\emph{Phys. Lett. B}
  {\bfseries 670} (2008) 84} [\href{https://arxiv.org/abs/0805.4346}{{\ttfamily
  0805.4346}}].

\bibitem{Albacete:2008ze}
J.L.~Albacete, Y.V.~Kovchegov and A.~Taliotis, \emph{{DIS on a Large Nucleus in
  AdS/CFT}}, \href{https://doi.org/10.1088/1126-6708/2008/07/074}{\emph{JHEP}
  {\bfseries 07} (2008) 074} [\href{https://arxiv.org/abs/0806.1484}{{\ttfamily
  0806.1484}}].

\bibitem{Penedones:2007ns}
J.~Penedones, \emph{{High Energy Scattering in the AdS/CFT Correspondence}},
  other thesis, 12, 2007, [\href{https://arxiv.org/abs/0712.0802}{{\ttfamily
  0712.0802}}].

\bibitem{BallonBayona:2007qr}
C.A.~Ballon~Bayona, H.~Boschi-Filho and N.R.F.~Braga, \emph{{Deep inelastic
  scattering from gauge string duality in the soft wall model}},
  \href{https://doi.org/10.1088/1126-6708/2008/03/064}{\emph{JHEP} {\bfseries
  03} (2008) 064} [\href{https://arxiv.org/abs/0711.0221}{{\ttfamily
  0711.0221}}].

\bibitem{Cornalba:2008sp}
L.~Cornalba and M.S.~Costa, \emph{{Saturation in Deep Inelastic Scattering from
  AdS/CFT}}, \href{https://doi.org/10.1103/PhysRevD.78.096010}{\emph{Phys. Rev.
  D} {\bfseries 78} (2008) 096010}
  [\href{https://arxiv.org/abs/0804.1562}{{\ttfamily 0804.1562}}].

\bibitem{Aharony:1999ti}
O.~Aharony, S.S.~Gubser, J.M.~Maldacena, H.~Ooguri and Y.~Oz, \emph{{Large N
  field theories, string theory and gravity}},
  \href{https://doi.org/10.1016/S0370-1573(99)00083-6}{\emph{Phys. Rept.}
  {\bfseries 323} (2000) 183}
  [\href{https://arxiv.org/abs/hep-th/9905111}{{\ttfamily hep-th/9905111}}].

\bibitem{Bianchi:2001de}
M.~Bianchi, D.Z.~Freedman and K.~Skenderis, \emph{{How to go with an RG flow}},
  \href{https://doi.org/10.1088/1126-6708/2001/08/041}{\emph{JHEP} {\bfseries
  08} (2001) 041} [\href{https://arxiv.org/abs/hep-th/0105276}{{\ttfamily
  hep-th/0105276}}].

\bibitem{Bianchi:2001kw}
M.~Bianchi, D.Z.~Freedman and K.~Skenderis, \emph{{Holographic
  renormalization}},
  \href{https://doi.org/10.1016/S0550-3213(02)00179-7}{\emph{Nucl. Phys. B}
  {\bfseries 631} (2002) 159}
  [\href{https://arxiv.org/abs/hep-th/0112119}{{\ttfamily hep-th/0112119}}].

\bibitem{Zou:2018eam}
L.~Zou and H.G.~Dosch, \emph{{A very Practical Guide to Light Front Holographic
  QCD}},  \href{https://arxiv.org/abs/1801.00607}{{\ttfamily 1801.00607}}.

\bibitem{Witten:1998zw}
E.~Witten, \emph{{Anti-de Sitter space, thermal phase transition, and
  confinement in gauge theories}},
  \href{https://doi.org/10.4310/ATMP.1998.v2.n3.a3}{\emph{Adv. Theor. Math.
  Phys.} {\bfseries 2} (1998) 505}
  [\href{https://arxiv.org/abs/hep-th/9803131}{{\ttfamily hep-th/9803131}}].

\bibitem{Brandhuber:1998er}
A.~Brandhuber, N.~Itzhaki, J.~Sonnenschein and S.~Yankielowicz, \emph{{Wilson
  loops, confinement, and phase transitions in large N gauge theories from
  supergravity}},
  \href{https://doi.org/10.1088/1126-6708/1998/06/001}{\emph{JHEP} {\bfseries
  06} (1998) 001} [\href{https://arxiv.org/abs/hep-th/9803263}{{\ttfamily
  hep-th/9803263}}].

\bibitem{Brower:2000rp}
R.C.~Brower, S.D.~Mathur and C.-I.~Tan, \emph{{Glueball spectrum for QCD from
  AdS supergravity duality}},
  \href{https://doi.org/10.1016/S0550-3213(00)00435-1}{\emph{Nucl. Phys.}
  {\bfseries B587} (2000) 249}
  [\href{https://arxiv.org/abs/hep-th/0003115}{{\ttfamily hep-th/0003115}}].

\bibitem{Brunner:2015oqa}
F.~Br{\"u}nner, D.~Parganlija and A.~Rebhan, \emph{{Glueball Decay Rates in the
  Witten-Sakai-Sugimoto Model}},
  \href{https://doi.org/10.1103/PhysRevD.91.106002}{\emph{Phys. Rev.}
  {\bfseries D91} (2015) 106002}
  [\href{https://arxiv.org/abs/1501.07906}{{\ttfamily 1501.07906}}].

\bibitem{Kinar:1998vq}
Y.~Kinar, E.~Schreiber and J.~Sonnenschein, \emph{{Q anti-Q potential from
  strings in curved space-time: Classical results}},
  \href{https://doi.org/10.1016/S0550-3213(99)00652-5}{\emph{Nucl. Phys.}
  {\bfseries B566} (2000) 103}
  [\href{https://arxiv.org/abs/hep-th/9811192}{{\ttfamily hep-th/9811192}}].

\bibitem{Gursoy:2007cb}
U.~Gursoy and E.~Kiritsis, \emph{{Exploring improved holographic theories for
  QCD: Part I}},
  \href{https://doi.org/10.1088/1126-6708/2008/02/032}{\emph{JHEP} {\bfseries
  02} (2008) 032} [\href{https://arxiv.org/abs/0707.1324}{{\ttfamily
  0707.1324}}].

\bibitem{Gursoy:2007er}
U.~Gursoy, E.~Kiritsis and F.~Nitti, \emph{{Exploring improved holographic
  theories for QCD: Part II}},
  \href{https://doi.org/10.1088/1126-6708/2008/02/019}{\emph{JHEP} {\bfseries
  02} (2008) 019} [\href{https://arxiv.org/abs/0707.1349}{{\ttfamily
  0707.1349}}].

\bibitem{Gursoy:2008za}
U.~Gursoy, E.~Kiritsis, L.~Mazzanti and F.~Nitti, \emph{{Holography and
  Thermodynamics of 5D Dilaton-gravity}},
  \href{https://doi.org/10.1088/1126-6708/2009/05/033}{\emph{JHEP} {\bfseries
  05} (2009) 033} [\href{https://arxiv.org/abs/0812.0792}{{\ttfamily
  0812.0792}}].

\bibitem{Karch:2002sh}
A.~Karch and E.~Katz, \emph{{Adding flavor to AdS / CFT}},
  \href{https://doi.org/10.1088/1126-6708/2002/06/043}{\emph{JHEP} {\bfseries
  06} (2002) 043} [\href{https://arxiv.org/abs/hep-th/0205236}{{\ttfamily
  hep-th/0205236}}].

\bibitem{Bianchi:2013gka}
M.~Bianchi, G.~Inverso, J.F.~Morales and D.~Ricci~Pacifici, \emph{{Unoriented
  Quivers with Flavour}},
  \href{https://doi.org/10.1007/JHEP01(2014)128}{\emph{JHEP} {\bfseries 01}
  (2014) 128} [\href{https://arxiv.org/abs/1307.0466}{{\ttfamily 1307.0466}}].

\bibitem{Addazi:2015rwa}
A.~Addazi and M.~Bianchi, \emph{{Un-oriented Quiver Theories for Majorana
  Neutrons}}, \href{https://doi.org/10.1007/JHEP07(2015)144}{\emph{JHEP}
  {\bfseries 07} (2015) 144}
  [\href{https://arxiv.org/abs/1502.01531}{{\ttfamily 1502.01531}}].

\bibitem{Addazi:2015hka}
A.~Addazi and M.~Bianchi, \emph{{Neutron Majorana mass from Exotic Instantons
  in a Pati-Salam model}},
  \href{https://doi.org/10.1007/JHEP06(2015)012}{\emph{JHEP} {\bfseries 06}
  (2015) 012} [\href{https://arxiv.org/abs/1502.08041}{{\ttfamily
  1502.08041}}].

\bibitem{Bianchi:2020fuk}
M.~Bianchi, D.~Bufalini, S.~Mancani and F.~Riccioni, \emph{{Mass deformations
  of unoriented quiver theories}},
  \href{https://doi.org/10.1007/JHEP07(2020)015}{\emph{JHEP} {\bfseries 07}
  (2020) 015} [\href{https://arxiv.org/abs/2003.09620}{{\ttfamily
  2003.09620}}].

\bibitem{Antinucci:2020yki}
A.~Antinucci, S.~Mancani and F.~Riccioni, \emph{{Infrared duality in unoriented
  Pseudo del Pezzo}},
  \href{https://doi.org/10.1016/j.physletb.2020.135902}{\emph{Phys. Lett. B}
  {\bfseries 811} (2020) 135902}
  [\href{https://arxiv.org/abs/2007.14749}{{\ttfamily 2007.14749}}].

\bibitem{Antinucci:2021edv}
A.~Antinucci, M.~Bianchi, S.~Mancani and F.~Riccioni, \emph{{Suspended Fixed
  Points}},  \href{https://arxiv.org/abs/2105.06195}{{\ttfamily 2105.06195}}.

\bibitem{Amariti:2021lhk}
A.~Amariti, M.~Fazzi, S.~Rota and A.~Segati, \emph{{Conformal S-dualities from
  O-planes}},  \href{https://arxiv.org/abs/2108.05397}{{\ttfamily 2108.05397}}.

\bibitem{Etxebarria:2021lmq}
I.n.G.~Etxebarria, B.~Heidenreich, M.~Lotito and A.K.~Sorout,
  \emph{{Deconfining $\mathcal{N}=2$ SCFTs, or the Art of Brane Bending}},
  \href{https://arxiv.org/abs/2111.08022}{{\ttfamily 2111.08022}}.

\bibitem{Kruczenski:2003be}
M.~Kruczenski, D.~Mateos, R.C.~Myers and D.J.~Winters, \emph{{Meson
  spectroscopy in AdS / CFT with flavor}},
  \href{https://doi.org/10.1088/1126-6708/2003/07/049}{\emph{JHEP} {\bfseries
  07} (2003) 049} [\href{https://arxiv.org/abs/hep-th/0304032}{{\ttfamily
  hep-th/0304032}}].

\bibitem{Burrington:2004id}
B.A.~Burrington, J.T.~Liu, L.A.~Pando~Zayas and D.~Vaman, \emph{{Holographic
  duals of flavored N=1 super Yang-mills: Beyond the probe approximation}},
  \href{https://doi.org/10.1088/1126-6708/2005/02/022}{\emph{JHEP} {\bfseries
  02} (2005) 022} [\href{https://arxiv.org/abs/hep-th/0406207}{{\ttfamily
  hep-th/0406207}}].

\bibitem{Sakai:2003wu}
T.~Sakai and J.~Sonnenschein, \emph{{Probing flavored mesons of confining gauge
  theories by supergravity}},
  \href{https://doi.org/10.1088/1126-6708/2003/09/047}{\emph{JHEP} {\bfseries
  09} (2003) 047} [\href{https://arxiv.org/abs/hep-th/0305049}{{\ttfamily
  hep-th/0305049}}].

\bibitem{Klebanov:2000hb}
I.R.~Klebanov and M.J.~Strassler, \emph{{Supergravity and a confining gauge
  theory: Duality cascades and chi SB resolution of naked singularities}},
  \href{https://doi.org/10.1088/1126-6708/2000/08/052}{\emph{JHEP} {\bfseries
  08} (2000) 052} [\href{https://arxiv.org/abs/hep-th/0007191}{{\ttfamily
  hep-th/0007191}}].

\bibitem{Dymarsky:2009cm}
A.~Dymarsky, S.~Kuperstein and J.~Sonnenschein, \emph{{Chiral Symmetry Breaking
  with non-SUSY D7-branes in ISD backgrounds}},
  \href{https://doi.org/10.1088/1126-6708/2009/08/005}{\emph{JHEP} {\bfseries
  08} (2009) 005} [\href{https://arxiv.org/abs/0904.0988}{{\ttfamily
  0904.0988}}].

\bibitem{Sakai:2005yt}
T.~Sakai and S.~Sugimoto, \emph{{More on a holographic dual of QCD}},
  \href{https://doi.org/10.1143/PTP.114.1083}{\emph{Prog. Theor. Phys.}
  {\bfseries 114} (2005) 1083}
  [\href{https://arxiv.org/abs/hep-th/0507073}{{\ttfamily hep-th/0507073}}].

\bibitem{Rebhan:2014rxa}
A.~Rebhan, \emph{{The Witten-Sakai-Sugimoto model: A brief review and some
  recent results}},
  \href{https://doi.org/10.1051/epjconf/20159502005}{\emph{EPJ Web Conf.}
  {\bfseries 95} (2015) 02005}
  [\href{https://arxiv.org/abs/1410.8858}{{\ttfamily 1410.8858}}].

\bibitem{Jarvinen:2011qe}
M.~Jarvinen and E.~Kiritsis, \emph{{Holographic Models for QCD in the Veneziano
  Limit}}, \href{https://doi.org/10.1007/JHEP03(2012)002}{\emph{JHEP}
  {\bfseries 03} (2012) 002} [\href{https://arxiv.org/abs/1112.1261}{{\ttfamily
  1112.1261}}].

\bibitem{Bianchi:2020cfc}
M.~Bianchi, D.~Consoli and P.~Di~Vecchia, \emph{{On the N-pion extension of the
  Lovelace-Shapiro model}},
  \href{https://doi.org/10.1007/JHEP03(2021)119}{\emph{JHEP} {\bfseries 03}
  (2021) 119} [\href{https://arxiv.org/abs/2002.05419}{{\ttfamily
  2002.05419}}].

\bibitem{Bianchi:2015vsa}
M.~Bianchi and D.~Consoli, \emph{{Simplifying one-loop amplitudes in
  superstring theory}},
  \href{https://doi.org/10.1007/JHEP01(2016)043}{\emph{JHEP} {\bfseries 01}
  (2016) 043} [\href{https://arxiv.org/abs/1508.00421}{{\ttfamily
  1508.00421}}].

\bibitem{Cappelli:2012cto}
A.~Cappelli, E.~Castellani, F.~Colomo and P.~Di~Vecchia, eds., \emph{{The birth
  of string theory}}, Cambridge Univ. Press, Cambridge, UK (5, 2012).

\bibitem{Ammon:2015wua}
M.~Ammon and J.~Erdmenger, \emph{{Gauge/gravity duality}: {Foundations and
  applications}}, Cambridge University Press, Cambridge (4, 2015).

\bibitem{Mintakevich:2008mm}
O.~Mintakevich and J.~Sonnenschein, \emph{{On the spectra of scalar mesons from
  HQCD models}},
  \href{https://doi.org/10.1088/1126-6708/2008/08/082}{\emph{JHEP} {\bfseries
  08} (2008) 082} [\href{https://arxiv.org/abs/0806.0152}{{\ttfamily
  0806.0152}}].

\bibitem{Kaplunovsky:2015zsa}
V.~Kaplunovsky, D.~Melnikov and J.~Sonnenschein, \emph{{Holographic Baryons and
  Instanton Crystals}},
  \href{https://doi.org/10.1142/S0217984915400527}{\emph{Mod. Phys. Lett. B}
  {\bfseries 29} (2015) 1540052}
  [\href{https://arxiv.org/abs/1501.04655}{{\ttfamily 1501.04655}}].

\bibitem{Armoni:2016nzm}
A.~Armoni and E.~Ireson, \emph{{Holographic Corrections to the Veneziano
  Amplitude}},
  \href{https://doi.org/10.1016/j.physletb.2017.05.079}{\emph{Phys. Lett. B}
  {\bfseries 771} (2017) 430}
  [\href{https://arxiv.org/abs/1607.04422}{{\ttfamily 1607.04422}}].

\bibitem{Armoni:2009vv}
A.~Armoni, A.~Giveon, D.~Israel and V.~Niarchos, \emph{{Brane Dynamics and 3D
  Seiberg Duality on the Domain Walls of 4D N=1 SYM}},
  \href{https://doi.org/10.1088/1126-6708/2009/07/061}{\emph{JHEP} {\bfseries
  07} (2009) 061} [\href{https://arxiv.org/abs/0905.3195}{{\ttfamily
  0905.3195}}].

\bibitem{Gross:1987ar}
D.J.~Gross and P.F.~Mende, \emph{{String Theory Beyond the Planck Scale}},
  \href{https://doi.org/10.1016/0550-3213(88)90390-2}{\emph{Nucl. Phys. B}
  {\bfseries 303} (1988) 407}.

\bibitem{Gross:1987kza}
D.J.~Gross and P.F.~Mende, \emph{{The High-Energy Behavior of String Scattering
  Amplitudes}}, \href{https://doi.org/10.1016/0370-2693(87)90355-8}{\emph{Phys.
  Lett. B} {\bfseries 197} (1987) 129}.

\bibitem{Mende:1989wt}
P.F.~Mende and H.~Ooguri, \emph{{Borel Summation of String Theory for Planck
  Scale Scattering}},
  \href{https://doi.org/10.1016/0550-3213(90)90202-O}{\emph{Nucl. Phys. B}
  {\bfseries 339} (1990) 641}.

\end{thebibliography}\endgroup
\end{document}